\documentclass{emulateapj}
\usepackage{natbib}
\usepackage{graphicx}
\graphicspath{{./paper_plots/sig_e_plots/unnorm/}{./paper_plots/sig_e_plots/frac/}{./paper_plots/slit_profiles/}{./paper_plots/fov/5001/}{./paper_plots/fov/5003/}{../fov/compressed/}{./paper_plots/sig_e_plots/vsre/}{./paper_plots/mass_enc/}{./paper_plots/mass_vs_sig/}{./paper_plots/anisotropy/}{./paper_plots/ltrans/}{./paper_plots/fov_subtract/}}

\newcommand{\kms}{\hbox{km s$^{-1}$}}

\newcounter{minirefcount}
\usepackage{amssymb}
\usepackage{amsmath}
\usepackage{wasysym}
\usepackage{color}
\usepackage[normalem]{ulem}

\newcommand{\mcmc}{M_{\mathrm{CMC}}}
\newcommand{\epscmc}{\epsilon_{\mathrm{CMC}}}
\newcommand{\vlos}{v_{\mathrm{los}}}
\newcommand{\siglos}{\sigma_{\mathrm{los}}}
\newcommand{\sigreav}{\sigma}
\newcommand{\sigoav}{\langle \sige \rangle_{\philon,i}}
\newcommand{\philon}{\Phi_{\mathrm{LON}}}
\newcommand{\sige}{\sigma}
\newcommand{\re}{R_{e}}
\newcommand{\mbh}{M_{\mathrm{BH}}}
\newcommand{\msigmarelation}{\log M_{\mathrm{BH}} = \alpha + \beta \log (\sige/200 \hspace{2 pt} \mathrm{km \hspace{2 pt} s^{-1}})}

\newcommand{\msol}{M_{\odot}}
\newcommand{\sigr}{\sigma_{R}}
\newcommand{\sigp}{\sigma_{\phi}}
\newcommand{\sigz}{\sigma_{z}}
\newcommand{\betap}{\beta_{\phi}}
\newcommand{\betaz}{\beta_{z}}
\newcommand{\msigma}{\mbh-\sige}

\begin{document}

\title{On The Offset of Barred Galaxies from the Black Hole $\msigma$ Relationship}

\author{Jonathan S. Brown\altaffilmark{1}}  
\author{Monica Valluri\altaffilmark{1}}
\author{Juntai Shen\altaffilmark{2}}
\author{Victor P. Debattista\altaffilmark{3}}

\altaffiltext{1}{Department of Astronomy, University of Michigan, Ann Arbor,
  MI 48109, USA; Email: brojonat@umich.edu, mvalluri@umich.edu;}  
\altaffiltext{2}{Key Laboratory for Research in Galaxies and Cosmology, Shanghai Astronomical Observatory, Chinese Academy of Sciences, 80 Nandan Road, Shanghai 200030, China;}
\altaffiltext{3}{Jeremiah Horrocks Institute, University of Central Lancashire, Preston, PR1 2HE, United Kingdom;}

\slugcomment{Draft Version of  \textit{\today}}

\begin{abstract}
We use collisionless $N$-body simulations to determine how the growth of a supermassive black hole (SMBH) influences the nuclear kinematics in both barred and unbarred galaxies. In the presence of a bar, the increase in the velocity dispersion $\sigma$ (within the effective radius) due to the growth of an SMBH is on average $\lesssim 10\%$, whereas the increase is only $\lesssim 4\%$ in an unbarred galaxy. In a barred galaxy, the increase results from a combination of three separate factors (a) orientation and inclination effects; (b) angular momentum transport by the bar that results in an increase in the central mass density; (c) an increase in the vertical and radial velocity anisotropy of stars in the vicinity of the SMBH. In contrast the growth of the SMBH in an unbarred galaxy causes the velocity distribution in the inner part of the nucleus to become less radially anisotropic. The increase in $\sigma$ following the growth of the SMBH is insensitive to a variation of a factor of 10 in the final mass of the SMBH, showing that it is the growth process rather than the actual SMBH mass that alters bar evolution in a way that increases $\sigma$. We argue that using an axisymmetric stellar dynamical modeling code to measure SMBH masses in barred galaxies could result in a slight overestimate of the derived $\mbh$, especially if a constant M/L ratio is assumed. We conclude that the growth of a black hole in the presence of a bar could result in an increase in $\sigma$ which is roughly of 4-8\% larger than the increase that occurs in an axisymmetric system. While the increase in $\sigma$ due to SMBH growth in a barred galaxy might  partially account for the claimed offset of barred galaxies  and pseudo bulges from the $\msigma$ relation obtained for elliptical galaxies and classical bulges in unbarred galaxies, it is inadequate to account for all of the offset.
\end{abstract}


\section{Introduction}
\label{sec:intro}

Over the past 20 years it has become increasingly evident that nearly all massive galaxies have a supermassive black hole (SMBH) residing at their centers \citep{Kormendy95,Magorrian98,Richstone98}. A growing sample of dynamically measured black hole masses has allowed for the development and refinement of important scaling relations between SMBHs and their host galaxies. Many scaling relations have been established, including those that relate the mass of the SMBH, (hereafter $\mbh$), to properties of the host spheroid/bulge/elliptical, e.g. spheroid mass $M_{\rm bul}$, bulge luminosity $L_{\rm bul}$ \citep{Kormendy95,Magorrian98,Richstone98,Marconi03,Haring04}, stellar velocity dispersion within the half-light radius $\sige$ \citep[the $\msigma$ relation][]{Ferrarese00,Gebhardt00,Tremaine02}, the circular velocity of the dark matter halo $v_{\rm circ}$ \citep{Ferrarese02}, the S\'{e}sic index of the bulge $n$ \citep{Graham07}, the number of globular clusters \citep{Burkert10,Harris11}, and even the spiral arm pitch angle of the galaxy \citep{Seigar08,Ringermacher09}. These scaling relations imply a strong coupling between the SMBH at a galaxy's center and the global properties of the galaxy itself.  A complete understanding of these scaling relations, and the causes of any deviations, will enable us to infer more accurately e.g. the masses of SMBH in distant galaxies where direct $\mbh$ measurements are not possible. Theoretical investigations of the physical causes of deviations from scaling relations can enhance our understanding of the growth and co-evolution of SMBHs and their host galaxies over cosmic time.

The tightest and most extensively studied of the SMBH scaling relations is the $\msigma$ relation, which takes the form $\msigmarelation$. Since the contemporaneous papers by \citet{Gebhardt00} and \citet{Ferrarese00} established values for the slope $\beta$ of the relation as  $3.75 \pm 0.3$ and $4.80 \pm 0.54$ respectively, there have been numerous revisions and recalculations of the slope, including $4.02 \pm 0.32$ \citep{Tremaine02}, $4.86 \pm 0.43$ \citep{Ferrarese05}, $4.24 \pm 0.41$ \citep{Gultekin09b}, $5.13\pm0.34$ \citep{Graham11} and most recently $5.64 \pm 0.32$ \citep{mcconnell_ma_13}. 

As the number of galaxies with measured $\mbh$ has grown, attempts have been made to examine whether the scaling relations are dependent on the morphological type of the host galaxies. Some recent studies have shown that barred galaxies may be offset from the $\msigma$ relationship obtained for unbarred galaxies \citep[e.g.,][]{Hu08,Graham08a,Graham08b,Graham09,Graham11}.  \citet{Graham09} found that if barred galaxies are excluded from the $\msigma$ relationship, the scatter in the relation drops from 0.47 dex to 0.41 dex. Furthermore, \citet{Graham11} showed that barred galaxies reside $\sim 0.30$ dex below the $\msigma$ relation defined by unbarred galaxies (classical bulges and elliptical galaxies), although both populations follow parallel scaling relations with $\beta \sim 5$. However,  in a study of the $\msigma$ relation for AGN, \citet{Xiao11} find that there is no significant difference in the slope $\beta$ for barred and unbarred AGN, but these authors do find a small offset between low-inclination and high-inclination disk galaxies (highly inclined galaxies have larger $\sigma$ at a given value of BH mass). A  study of the $\msigma$ relation in $\sim$150 galaxies (including $\sim 100$ upper limits) found no offset between barred and unbarred galaxies \citep{Beifiori12}. 
\citet{Greene10} found that $\mbh$ values measured in a sample of late-type Seyfert II galaxies were about a factor of two smaller than $\mbh$ values predicted from the observed $\sige$ using the standard $\msigma$ relationship. This is consistent with a  recent examination of the $\msigma$ relationship for early type galaxies vs. late type galaxies \citep{mcconnell_ma_13} which shows that both types have consistent slopes ($\beta = 5.2 \pm 0.36$ and $\beta = 5.06 \pm 1.16$ respectively), but the late-type galaxies have a significantly {\em lower} zero-point $\alpha.$ 

\citet{Graham11} find that the offset of barred galaxies from the $\msigma$-relationship for unbarred galaxies is 0.3~dex in $\mbh$, assuming a slope of $\beta \simeq 5$, this corresponds a rightward offset of 0.06~dex in $\sige$. This implies that on average, the stellar velocity dispersion of barred disk galaxies is $\sim 15 \%$\footnote{$\beta=4$ would imply an increase in $\sigma$ of $\sim$ 19\%} higher than that of unbarred disk  galaxies. Recently \citet{Hartmann13} re-evaluated the offset of barred galaxies with classical bulges from the $\msigma$ relation for unbarred galaxies with classical bulges and find an offset of 0.2~dex (and a scatter of 0.19~dex). However they  ind that barred galaxies with pseudo-bulges are offset by 0.4~dex from the $\msigma$ relationship of unbarred classical bulges. The intermediate value of 0.3~dex found by \citet{Graham11} probably results from their inclusion of barred galaxies with pseudo bulges. This larger offset for pseudo bulges is consistent with the finding of \citet{Kormendy_bender_11} that pseudo-bulges do not follow the $\msigma$ relation defined by elliptical galaxies and classical bulges.

 \citet{Graham11} offer several possible explanations for systematically large observed $\sigma$ for the barred sample. These include viewing angle --  the orientation of the bar to the line-of-sight, and the inclination of the disk (which can cause contamination of $\sigma$ by disk particles), and the presence of nuclear star clusters.  Also \citet{Graham08a}  examined the possibility that the offset of barred galaxies could be the consequence of their having undermassive SMBHs as opposed to their having systematically higher velocity dispersions than their unbarred counterparts. He argued that since barred galaxies are not offset from the $\mbh$--$L$ relation, anemic SMBHs are not to blame. 
  
 \citet{Hartmann13} use $N$-body simulations to examine the effects of bar formation and evolution on the observed $\sigma$ in bar-unstable disk galaxies with classical bulges. They analyze a set of 25 disk+bulge simulations both before and after bar formation. It is well known that bar formation in an initially cold disk followed by bar buckling can lead to a redistribution of angular momentum and kinetic energy that results in the heating of the disk \citep[e.g][]{Hohl71,Raha91} and the formation of a boxy, peanut-shaped bulge. The simulations examined by \citet{Hartmann13} do not include the growth of a point mass representing an SMBH. Rather they assume that each bulge contains an SMBH whose mass is set by the $\msigma$ relationship and $\mbh$ does not change as the bar evolves. 

In this paper we examine via $N$-body simulations whether the claimed offset of bars from the $\msigma$ relation could be a consequence of the effects of the dynamical evolution of a  bar resulting from the growth of a central black hole on the observed value of $\sige$. We also discuss how stellar dynamical measurements of $\mbh$ may be affected.  We analyze a set of $N$-body simulations of barred galaxies (and unbarred counterparts constructed from them) both with and without classical bulges. Central mass concentrations (CMC) representing SMBHs are grown adiabatically in each of our disk galaxies, and the dynamical response of the barred or unbarred disk galaxy is examined.  

Although it has long been thought that the feeding of a central AGN and the resulting growth of the  central black hole could be a consequence of the evolution of a bar and gas transport by it \citep{Simkin80} the evidence for a direct connection between bars and AGN growth remains elusive \citep[e.g.][]{Oh12}. The study by \citet{Hartmann13} and the one presented here are complementary in that they span two extremes of the  range of possibilities: \citet{Hartmann13} explore the effects of {\it bar formation and evolution  on bulges assumed to have pre-existing SMBHs}, while we examine the effect of the {\it  adiabatic growth of a SMBH on a pre-existing bar}. Reality probably lies somewhere in between these possibilities.

In Section 2 we describe the set up for the $N$-body simulations, in Section 3 we describe the analysis of these simulations, and in Section 4 we present the results of our analysis of the dynamical effects of bars and CMCs on observed 2D and 1D nuclear kinematics,  aperture dispersion, and velocity anisotropy. In Section 5 we summarize our results and in Section 6 we discuss their implications to our understanding of the co-evolution of galaxies and their SMBHs. 

\section{Simulations}
\label{sec:mod_set}

Our disk models, central mass concentration, and dark halo models are almost identical to those presented in \citet{Shen04}. We refer the reader  to this paper (and to references therein) for a more detailed description of the simulations. What follows is a brief description of each of the components of the simulations.

As is standard for such simulations the units used are, $G=M_d=R_d=1$ where G is Newton's gravitational constant, $M_d$ is the mass of the disk, and $R_d$ is the disk scale length. Dimensional arguments give a unit of time of $t_{\mathrm{dyn}} =(R_d^3/GM_d)^{1/2}$. We describe the initial configuration of the model in these units. Physically relevant scalings can be obtained by choosing observationally motivated values for $M_d$ and $R_d$. In this paper we adopt $M_d = 5 \times 10^{10} M_{\odot}$ and $R_d = 3$~kpc, which corresponds to a unit of time $t_{\mathrm{dyn}} \sim$ 11 Myr.  In all the figures and analysis that follows velocities are given in units of km s$^{-1}$ and distances in units of kpc, using this conversion.

We started with two types of initial conditions: one consisting of a pure disk (\S~\ref{sec:disk}) embedded in a static halo (\S~\ref{sec:halo}), and the second that also contains a spheroidal central distribution representing a classical bulge (\S~\ref{sec:bulge}).   Each set of initial conditions (at time $t_0$) was evolved until a time $t_1=700$ ($t_1=400$) for the pure disk (disk+bulge) simulations respectively.  During the time $t_0$ to $t_1$ the disks became bar unstable and the bars underwent buckling. At $t_1$ the bars in both simulations have reached a nearly steady state and have bulges which show the peanut shape characteristic of the buckling instability. Additionally,  the model with a pure disk has a boxy (pseudo) bulge, while in the model with a disk + classical bulge, it has a more oval shape. 

From each of the simulations at $t_1$ we constructed an unbarred ``control disk galaxy'' by repositioning each particle in the simulation at a randomly selected azimuthal angle $\phi$ while keeping their radius and vertical displacement from the disk plane fixed. The two resulting ``scrambled disks'' have the same radially averaged mass and velocity distributions as the two barred galaxies and enable us to compare and contrast the {\em dynamical} effects of the growth of an SMBH on bar, bulge, and disk particles. An important consequence of the ``scrambling'' process is that our unbarred (axisymmetric) disks are too hot to be able to subsequently form a bar, although they do form weak spirals, which produce slightly non-axisymmetric features following the growth of an SMBH.

 We grow a central mass concentration (CMC) representing an  SMBH  with two possible final masses ($\mcmc = 10^8\msol$ and $\mcmc = 10^7\msol$) in each of the above 4 simulations (see \S~\ref{sec:CMC} for details). The CMCs are grown adiabatically starting at an initial time $t_1$ and evolved until $t_2=1200$ ($t_2=900$) for the pure disk (disk+bulge) simulations respectively.  At $t_2$ the transients due to the changing CMC potential have  dissipated and the  simulations are examined and compared with those at $t_1$. 

Each simulation is then examine at two different times $t_1$ and $t_2$.  Thus in total we examine 2 snapshots each of 8 different simulations. In the figures that follow light colors (pink/cyan) are representative of simulations with $\mcmc = 10^7 \msol$, while dark colors (blue/red) show results for $\mcmc = 10^8 \msol$. Below we list the symbols/line styles used to denote each snapshot in the figures:\\
(1) a pure disk with a bar at $t_1$: open blue/cyan squares (denoting  a boxy-bulge) connected by dashed blue/cyan lines;\\
(2) a scrambled version of (1): open red/pink squares connected by dashed red/pink lines;\\
(3) model (1) after a  CMC was adiabatically grown: open blue/cyan squares connected by solid blue/cyan line;\\
(4) model (2) after a  CMC was adiabatically grown: open red/pink squares connected by solid red/pink lines;\\
(5)  disk+bulge with a bar at $t_1$: filled blue/cyan dots connected by dashed blue/cyan lines;\\
(6) a scrambled version of (5): filled red/pink dots connected by dashed red/pink lines;\\
(7) model (5) after a CMC was adiabatically grown: filled blue/cyan dots  (denoting a classical bulge) connected by solid blue/cyan lines;\\
(8) model (6) after a CMC was adiabatically grown: filled red/pink dots  connected by solid red/pink lines;\\

The set up of initial conditions for  particles  in each component of the disk galaxies used in our simulations and the  growth of the point mass are described in greater detail below.

\subsection{Disk Model}
\label{sec:disk}

The disk component is an evolved Kuz'min-Toomre (K-T) disk with the following surface density distribution
\begin{equation}
\Sigma(R) = \frac{M_d}{2\pi R_d^2}\left(1+\frac{R^2}{R_d^2}\right)^{-3/2}
\end{equation}
where $R$ is the radial distance from the axis of rotation and $R_d$ is the disk scale length. The disk is spread vertically as an isothermal sheet and truncated at $R = 5R_d$. Particles are drawn from a distribution function which yields a Toomre Q $\simeq$ 1.5.  The resulting structure is unstable to bar formation \citep{Athanassoula86}. The bar forms, and is vertically thickened via the buckling instability, resulting in a stable bar \citep{Toomre66, Raha91, Sellwood93}.

\subsection{Halo}
\label{sec:halo}

We choose a dark matter (DM) halo with the well known logarithmic potential
\begin{equation}
\Phi_{\mathrm{halo}}(r) = \frac{V^{2}_{0}}{2}\ln\left(1+\frac{r^2}{c^2}\right)
\end{equation}
which yields a flat circular velocity when $r \gg c$, where $c$ is the core radius \citep{Binney08}. We choose $c = 30 R_d = 90$~kpc, and $V_0 = 0.7(GM_d/R_d)^{1/2} = 187$~\kms. Since we use a rigid halo as opposed to a live halo, the halo in our simulations cannot exchange energy or angular momentum with the disk and/or bulge particles. \citet{Shen04} found that replacing their rigid logarithmic halo  with a live one resulted in little change to the evolution of the bar in their simulations. In these simulations the central region of the halo is shallow, preventing the halo from affecting the evolution of the angular momentum significantly. However \citet{athanassoula_lambert_dehnen_05} found in their simulations with live halos that the survival of the bar depended quite strongly on the density profile of the dark matter halo. They found that for  a CMC of the same mass, a  bar in a  DM halo with a shallow central central cusp is more easily destroyed than a bar in a DM halo with a steeply rising DM cusp.  In this paper we will assume only a rigid logarithmic halo with a core. We address the effect of this assumption on our results in \S~\ref{sec:discussion}. 

\subsection{Bulge Component}
\label{sec:bulge}

In the disk+bulge simulations, the bulge component has a mass of 0.15$M_d$ and is initially truncated at a radius of  {0.9}$R_d$. The two component system is constructed using a method first proposed by \citet{Prendergast70}, used in \citet{Raha91}, and described in \citet{Jarvis85} and Appendix A of \citet{Debattista00}. Using the integrals of motion $(E,J_z)$, a distribution function $f(E,J_z)$ is chosen that corresponds to a King model \citep{King66} with some net rotation \citep{Jarvis85}. Integrating the distribution function over velocity yields a density $\rho(R,z)$. The density is converted to a mass, which is added to that of a smooth disk component. The potential due to this new mass distribution is computed, yielding a new distribution function. This process is iterated until convergence.

\subsection{Central Mass Concentration}
\label{sec:CMC}

The central mass concentration (CMC) representing a SMBH is modeled as a Plummer sphere with potential of the form
\begin{equation}
\Phi_{\mathrm{CMC}}(r) = -\frac{GM_{\mathrm{CMC}}(t)}{\sqrt{r^2+\epscmc}}
\end{equation}
where $\epscmc$ is the softening length. The softening length corresponds to the compactness of the CMC. A large value of $\epscmc$ is representative of a relatively diffuse CMC (e.g. molecular gas clouds or a nuclear star cluster), whereas a small value represents a relatively compact (hard) CMC. \citet{Shen04} showed that the effect of a very compact CMC is much greater than that of a softer CMC. Here we set $\epscmc = 0.001R_d$ (corresponding to a length scale of a few parsecs)  since we wish to assess the stronger effect of its growth on the observable kinematics.

In half of our simulations, we choose a final $\mcmc$ of 0.2\% $M_{d}$ which for our choice of physical units corresponds to $10^8~\msol$. Note that this CMC is  a factor of 6.5 more massive than the SMBH mass predicted from scaling relation $\mbh \lesssim 0.002M_{bulge}$ \citep{Haring04}. For this reason, we also carry out an investigation of the effect of a CMC with 10 times smaller mass ($\mcmc = 10^7 \msol$) and show that the effects of this smaller black hole on the stellar velocity dispersion are similar to those resulting from the $\mcmc =10^8 \msol$. More importantly, the the fractional difference in $\sige$ between the barred and axisymmetric models is nearly independent of $\mcmc$. To remind readers that the central point mass in some of our simulations are somewhat overmassive we will henceforth refer to it as a CMC rather than a SMBH.

We adopt the definition for a black hole's ``sphere of influence'', $r_s$, as the radius within which the mass of stars is equal to the mass of the black hole. For $\mcmc = 10^8 \msol$ the sphere of influence $r_s = 0.17 \pm 0.078$~kpc. Since $r_s$ is directly proportional to the mass of the CMC, it is about a factor of 10 less for $\mcmc = 10^7 \msol$, which would make $r_s$ much smaller than the particle softening and therefore  unresolvable by our current simulations. Nevertheless, we will show that despite the factor of 10 difference in final masses of the two CMC, they both affects on the observed values of $\sigma$ in qualitatively similar ways and differing quantitatively by at most a few percent -- a difference that is unlikely to be observationally detectable.

The CMC is grown adiabatically on a timescale which is much longer than the orbital period of stars near the disk center. $\mcmc$ is a function of time given by
\begin{equation}
\mcmc(\tau) =\begin{cases}
0 & \tau < 0\\
\mcmc \sin^2(\pi \tau / 2) & 0 \leq \tau \leq 1\\
\mcmc & \tau > 1
\end{cases}
\end{equation}
where $\tau \equiv (t-t_{\mathrm{CMC}})/t_{grow}$ for a CMC which began growing at $t_{\mathrm{CMC}}$. We increase $\mcmc$ over $t_{\mathrm{grow}} = 50$~dynamical times.

\begin{deluxetable*}{lcccc}
\tablecaption{Summary of Model Setup}
\tablecolumns{8}
\tablehead{\colhead{Parameter} & \colhead{Disk} &\colhead{Disk+Bulge}}\\
\startdata
\cutinhead{Numerical Parameters}
Number of particles ............................................................................................................................ & 2.8 $\times 10^6$ & 1.15 $\times10^6$ \\
Grid size ($R$,$\phi$,$z$)..................................................................................................................................& $55 \times 64 \times 375$ & $58 \times 64 \times 375$ \\
Vertical plane spacing..........................................................................................................................& 0.02 & 0.01 \\
Grid boundaries ($R$, $z$).........................................................................................................................& (20.0, $\pm$ 3.74) & (26.8, $\pm$ 3.74)\\
Particle softening length.......................................................................................................................& 0.02 & 0.01 \\
Time step $\Delta t_0$ without CMC...............................................................................................................& 0.04 & 0.04 \\
Time step $\Delta t_0$ with CMC....................................................................................................................& 0.01 & 0.01 \\
Number of guard shells\tablenotemark{a}.......................................................................................................................& 9 & 9 \\
Outermost guard radius $r_{\mathrm{max}}$..............................................................................................................& 0.127 & 0.127 \\
Innermost guard radius $r_{\mathrm{min}}$................................................................................................................& 0.008 & 0.008 \\
Smallest time step................................................................................................................................ & $t_{step}/2^9$ & $t_{step}/2^9$ \\
\cutinhead{Initial Disk}
Toomre Q............................................................................................................................................. & 1.5 & 1.2 \\
RMS vertical thickness.........................................................................................................................& 0.3 & 0.5 \\
Truncation radius.................................................................................................................................& 5 & 5 \\
\cutinhead{Fixed Halo}
$V_0$......................................................................................................................................................... & 0.7 & 0.8 \\
Core radius $c$........................................................................................................................................& 30 & 8 \\
\cutinhead{Bulge}
Mass.....................................................................................................................................................& \nodata & 0.15 \\
Truncation radius.................................................................................................................................& \nodata & 0.9 \\
\cutinhead{CMC}
 $\mcmc$ (1)...................................................................................................................................... & $0.002$ & $0.002$ \\
$\mcmc$ (2)...................................................................................................................................... & $0.0002$ & $0.0002$ \\
Softening length $\epscmc$......................................................................................................................... & $0.001$ & $0.001$ \\
Growth time $t_{grow}$............................................................................................................................... &  $50$ & $50$ \\
\tablenotetext{a}{See Appendix of \citet{Shen04} for guard shell details}
\enddata
\label{tab:sims}
\end{deluxetable*}

\subsection{Numerical Methods}
\label{sec:nummethod}

The simulations use a three-dimensional, cylindrical, polar grid--based $N$-body code described in \citet{Sellwood97}. The gravitational field at a distance $d$ from a particle is given by a Plummer sphere $\Phi(d) = -G/(d^2+\epsilon^2)^{1/2}$. We use a constant particle softening length, $\epsilon = 0.02R_d$ in all of our simulations. See Table~\ref{tab:sims} for the full set of numerical parameters.

Due to the differing time scales associated with each particle, the simulation is divided into 4 spherical zones and different time steps are used in each zone, with the minimum timestep of $0.01/128$ \citep[for details see,][]{Shen04}. Additionally the ``guard-shell'' scheme described in detail in the Appendix of \citet{Shen04} (the CMC is enclosed by a number of spherical regions with successively shorter time steps as $R$ decreases) helps ensure accurate orbit integrations in areas where particles are subjected to relatively strong accelerations.

\section{Analysis of Simulations}
\label{sec:analysis}

For the analyses of the simulations we constructed two dimensional kinematic maps of each of the snapshots to represent the ``observable'' kinematics in 2 dimensional  ``integral field''  maps. Our main goals in this paper are  (a) to examine the dependence of $\sige$, the velocity dispersion within the half-light radius, on viewing angle (disk inclination and  angle of the bar to the line-of-nodes), and the presence or absence of a bar, bulge, or CMC; (b) to examine how the stellar nuclear kinematical quantities (that are normally used to measure the dynamical mass of the SMBH) differ between the barred and the unbarred systems.

To address the first goal we use the kinematic maps to compute $\sige$ for each of our simulations for a variety of viewing angles, from assumed values of $\re$.  In Section~\ref{sec:measure_mbh} we use ``difference maps'' representing the difference between the kinematic maps of barred and unbarred systems to examine the effects of bar dynamics on nuclear stellar kinematics.  We describe the computation of the kinematic maps and  $\sige$ below.

\subsection{Two Dimensional Kinematic Maps}
\label{sec:slit_extraction}

Our analysis begins by ``observing'' each snapshot at a specific angle of inclination of the disk to the line of sight, $i$, and the angle formed by the bar (if present) to the line of nodes, $\philon$.

Due to our focus on the nuclear region of the models, we restrict our field of view of the simulations to $\pm$10.5~kpc (and $\pm$ 7.5 kpc) in the $x$ and $y$ directions for the disk-only (and disk+bulge) simulations respectively. We binned all the particles that fall within this projected rectangular region on a 300$\times$300 Cartesian grid corresponding to a pixel size of 0.07$\times$0.07~kpc in the pure disk models (and pixels of 0.05$\times$0.05~kpc in the disk+bulge models). This is roughly equal to the particle softening length. We then adaptively bin the square pixels to maintain a minimum $S/N \equiv \sqrt{N} \geq 50$ using the Voronoi binning scheme outlined in \citet{Cappellari03}\footnote{We used M. Cappellari's IDL Voronoi binning routine available at http://www-astro.physics.ox.ac.uk/$\sim$mxc/idl/}. Our choice of pixel size and $S/N$ was a compromise between maintaining computational economy and attempting to resolve the sphere of influence $r_s\sim 0.17$~kpc of the $\mcmc = 10^8 \msol$.\footnote{recall that $r_s$ for the  $\mcmc = 10^7 \msol$ is not resolved by our simulations}.  We found that the resulting kinematics were relatively insensitive to our choice of pixel size and $S/N$ threshold, given a $S/N \gtrsim 30$. On average each Voronoi bin is composed of $\sim 300$ pixels, with the smallest and largest Voronoi bins containing 3 and 767 pixels respectively. Inside $R \sim 2$~kpc, individual pixels are comparable to the size of the Voronoi bins; outside of $R \sim 2$~kpc, the Voronoi bins are considerably larger than a single pixel.

We construct line-of-sight velocity distributions (LOSVDs) from all particles that fall within a Voronoi bin. Since the LOSVDs of such systems generally depart from pure Gaussian shapes, following the standard practice  we parametrized the LOSVD within each Voronoi bin using a Gauss-Hermite expansion \citep{VDM93,gerhard_93} and define $\vlos$ as the mean line-of-sight velocity, $\siglos$ as the line-of-sight velocity dispersion, and describe the asymmetric and symmetric  departures from a Gaussian LOSVD  by the Hermite coefficients $h_3$, $h_5$ and $h_4$,  $h_6$ respectively.  The parameters characterizing the LOSVD in each Voronoi bin were obtained with using the MPFIT procedure implemented in IDL \citep{Markwardt09} to simultaneously fit $\gamma$, $\vlos$ $\siglos$, $h_3$, $h_4$, $h_5$, and $h_6$.

Due to the anisotropic velocity distribution inherent to barred galaxies, both the inclination of the disk $i$ and the angle made by the bar to the line-of-nodes\footnote{Here we take the line-of-nodes to be the intersection of the disk plane to the plane of the sky and it is along the $x-$axis in our images.} $\philon$ are likely to alter the measured nuclear kinematics.  

\begin{figure*}

\centering{\includegraphics[scale=1.,width=1.\textwidth,trim=0.pt 0.pt 0.pt 300.pt,clip]{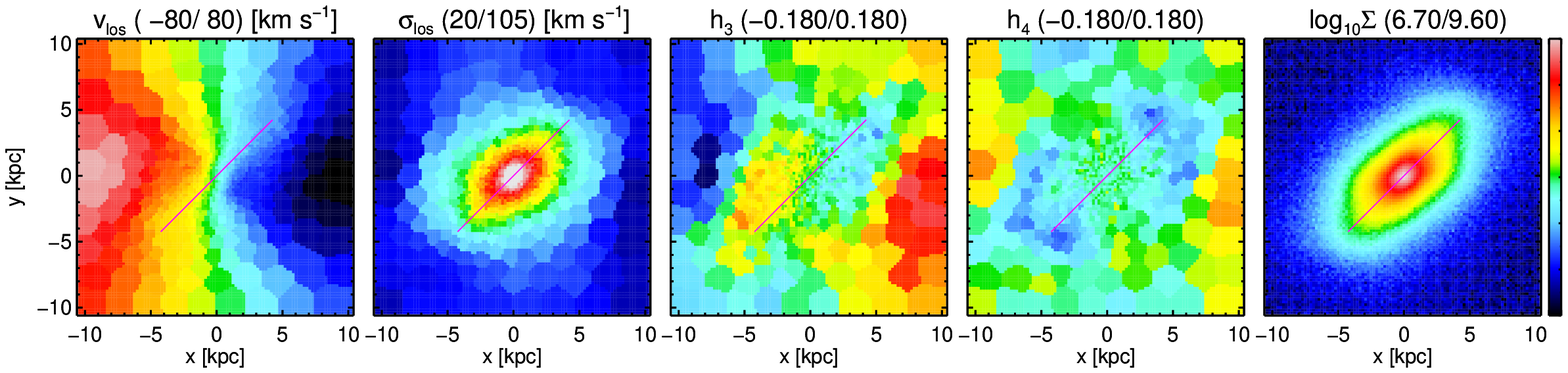}}

\centering{\includegraphics[scale=1.,width=\textwidth,trim=0.pt 0.pt 0.pt 365.pt,clip]{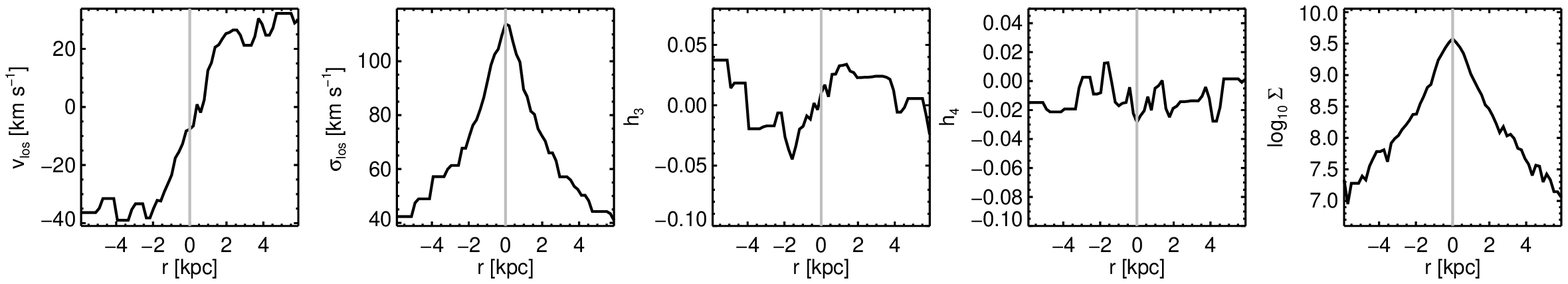}}

\caption{Top: 2-D kinematics ($\vlos, \siglos, h_3, h_4$) and surface brightness ($\log_{10}\Sigma$) for a 10.5~kpc $\times$~10.5~kpc field-of-view for the disk-only simulation with a bar, after the growth of the $ 10^8 \msol$ CMC. The quantities in parenthesis above each panel give the maximum (light red)/minimum (dark blue) of the quantity being plotted in that panel, with contours linearly spaced. The viewing angle is such that $i = 45^\circ$ and $\philon = 45^\circ$. The red line represents the slit used to extract the kinematics. The slit is oriented approximately along the bar passing through the center of the model. Bottom: The corresponding kinematics along the slit for each of the 4 kinematic parameters and surface brightness.}
\label{fig:50012d_and_slit}

\end{figure*}

Figure~\ref{fig:50012d_and_slit}~(top) shows the two dimensional kinematics fields (from left to right: $\vlos, \siglos, h_3, h_4$ and projected surface brightness $\log_{10}\Sigma$) for $i = 45^\circ$ and $\philon = 45^\circ$ for the disk-only simulation with a bar after the growth of the $\mcmc = 10^8 \msol$. The bottom panel shows the kinematics that would be observed along the artificial ``slit'' oriented along the major-axis of the bar (shown as a red line in the top panels).  For each rectangular  ``aperture'' along the slit, we average the kinematics of the bins which fall within that aperture. While we don't weight the bins according to the area of the slit they occupy (i.e. bins which fall only partially within a slit aperture are given the same weight as those which fall entirely within the aperture), we find that a more careful treatment  of apertures with partial overlap accounted for does not produce noticeable differences in the resulting slit profiles. In these figures we use a slit of length $-6 \leq r \leq 6$~kpc and width of 0.075 kpc (a factor of a few smaller than the sphere-of-influence of 0.17~kpc). 
 
\begin{figure*}

\centering{\includegraphics[scale=1.,width=1.\textwidth,trim=0.pt 0.pt 0.pt 250.pt,clip]{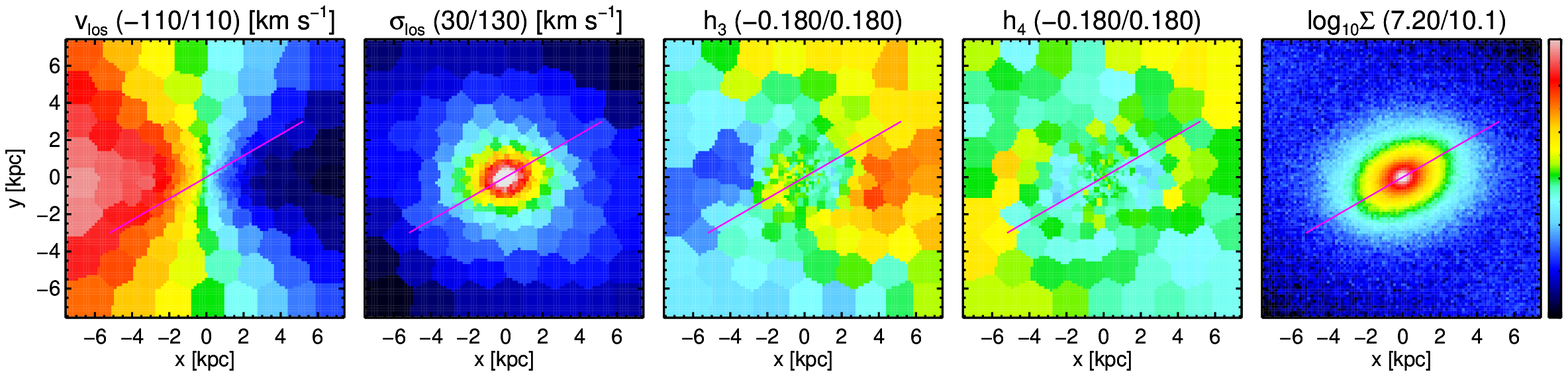}}

\centering{\includegraphics[scale=1.,width=\textwidth,trim=0.pt 0.pt 0.pt 295.pt,clip]{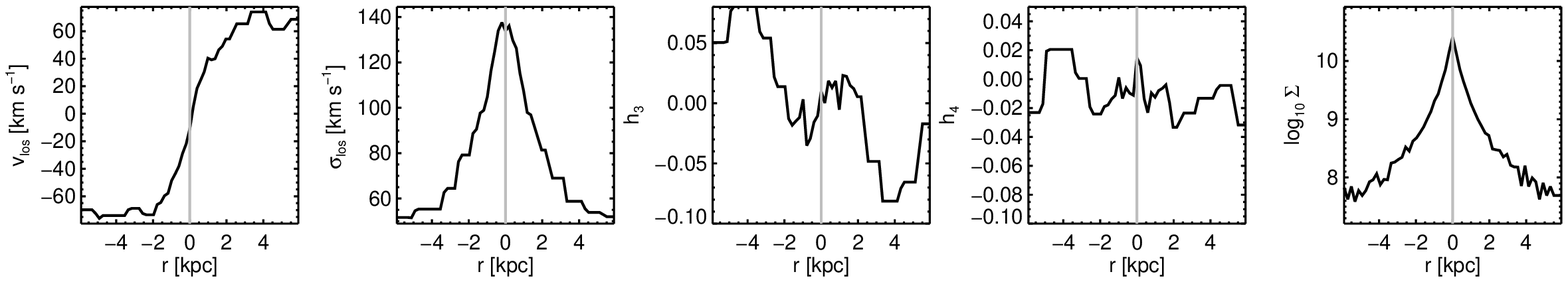}}

\caption{Same as Figure~\ref{fig:50012d_and_slit}, but for the 7.5~kpc $\times 7.5$~kpc field-of-view for the disk+bulge simulation with a bar, after the growth of the CMC.}

\label{fig:5003_2d_and_slit}

\end{figure*}

Similarly,  Figure~\ref{fig:5003_2d_and_slit} shows 2D kinematics (top) and slit-kinematics (bottom) (for $i = 45^\circ$ and $\philon = 45^\circ$) for the snapshot of the disk+bulge simulation with a bar after the growth of the a CMC with $\mcmc = 10^8 \msol$.  We note that in both the disk-only and disk+bulge simulations the $\vlos$ fields show a slight kinematic twist that is characteristic of triaxial systems and the rotational axis of symmetry is misaligned with the minor axis of the bar. In axisymmetric systems, $h_3$ is generally anticorrelated with $\vlos$, however in the region where the bar dominates $h_3$ tends to be correlated with $\vlos$  \citep{Bureau05}. This is indeed what we observe in both the disk-only and disk+bulge barred simulations, even in the presence of a CMC. Finally we observe the regions of negative $h_4$ that are characteristic of bars that have buckled and are then viewed face-on \citep{Debattista05}. In the model with the classical bulge (Fig.~\ref{fig:5003_2d_and_slit})  the bar is weaker than in Figure~\ref{fig:50012d_and_slit} however the kinematic twist in $\vlos$, the correlation between $h_3$ and $\vlos$, and the mis-alignment of the short-axis of the central oval and the rotation axis are tell-tale signs of the presence of a bar. 
 
\subsection{Computing $\sige$}
\label{sec:comp_sige}

The method and aperture used to define $\sige$ is a historically contentious issue \citep[e.g.][]{merritt_ferrarese_01,Tremaine02}. Here, we closely follow the observational definition of $\sige$ as the luminosity weighted RMS velocity within the projected half-light radius $\re$:
\begin{equation}
\sigma^2 = \frac{\int^{\re}_{0} I(R) (\sigma_{\mathrm{los}}^2+\overline{v}_{\mathrm{los}}^2) dR}{\int^{\re}_{0} I(R) dR,}
\label{eq:lum_sig}
\end{equation}
where $I(R)$ is the luminosity distribution of the bulge as a function of projected radius $R$, and $\sigma_{\mathrm{los}}$ and $\overline{v}_{\mathrm{los}}$ are the line-of-sight velocity dispersion and mean line-of-sight velocity respectively.
For our simulations, we assume that all particles are stars of the same type, that there is no dust, and that the stars have a constant mass-to-light ratio (i.e. $M/L = 1$). We then define a circular aperture of radius $\re$ that we project onto the field of view. We then convert the integral into a sum and compute $\sige$ as,
\begin{equation}
\sigma^2 = \frac{\sum\limits_{R_i \leqslant \re} m_i(\sigma_{i,los}^2+\overline{v}_{i,los}^2)}{\sum\limits_{R_i \leqslant \re} m_i,}
\label{eq:sum_sig}
\end{equation}
where the sum is over the cells on the 300$\times$300 grid which fall within $\re$, and $R_i$, $m_i$, are the projected distance from the center, mass, mean velocity and velocity dispersion of the $i$th cell respectively.  Note that this approach allows us to mimic what is done in IFU observations with a fixed pixel-scale.  

Since the orientation of the bar to the line-of-nodes as well as the inclination of the disk to the line-of-sight can alter $\sige$, we measured this quantity using Equation~\ref{eq:sum_sig} for 9 different orientations, as follows. With $i$ fixed at $45^\circ$ we varied the orientation of the bar so that $\philon =$ $0^\circ$, $30^\circ$, $45^\circ$, $60^\circ$, and $75^\circ$. We obtain 4 additional measurements with $\philon$ fixed at $45^\circ$ and inclination of the disk varied so that $i = 0^\circ$, $30^\circ$, $60^\circ$, and $75^\circ$.

Since a classical bulge is only present in half of the simulations,  $\re$ cannot be defined in a uniform way for all our simulations. Noting that when a bulge \emph{is} present, its truncation radius is $0.90R_d$ (2.7~kpc), we computed the mass within this radius (including the mass of disk particles interior to the truncation radius) and then (assuming that mass follows light with constant M/L) we compute the half-mass radius $r_{1/2}= 0.367R_d$ = 1.1~kpc\footnote{The half-mass radius is computed in cylindrical coordinates to be between 1.0-1.1~kpc, and slightly larger ($\sim 1.1-1.2$) when computed in spherical coordinates for the barred disk+bulge case at $t_1$ and $t_2$ respectively.}. We note that \citet{Hartmann13} show that, for their sample of simulations, the values of $\sige$ obtained using $\re/8$ are consistent with those obtained using $\re$.  We also tried four other possible values for $\re: $ 0.04, 0.08, 0.16, and $0.30R_d$ which correspond to values of 0.12, 0.24, 0.48, and 0.9~kpc respectively.
 
\begin{figure}

\centering{\includegraphics[scale=1.,width=.5\textwidth,trim=0.pt 0.pt 0.pt 0.pt,clip]{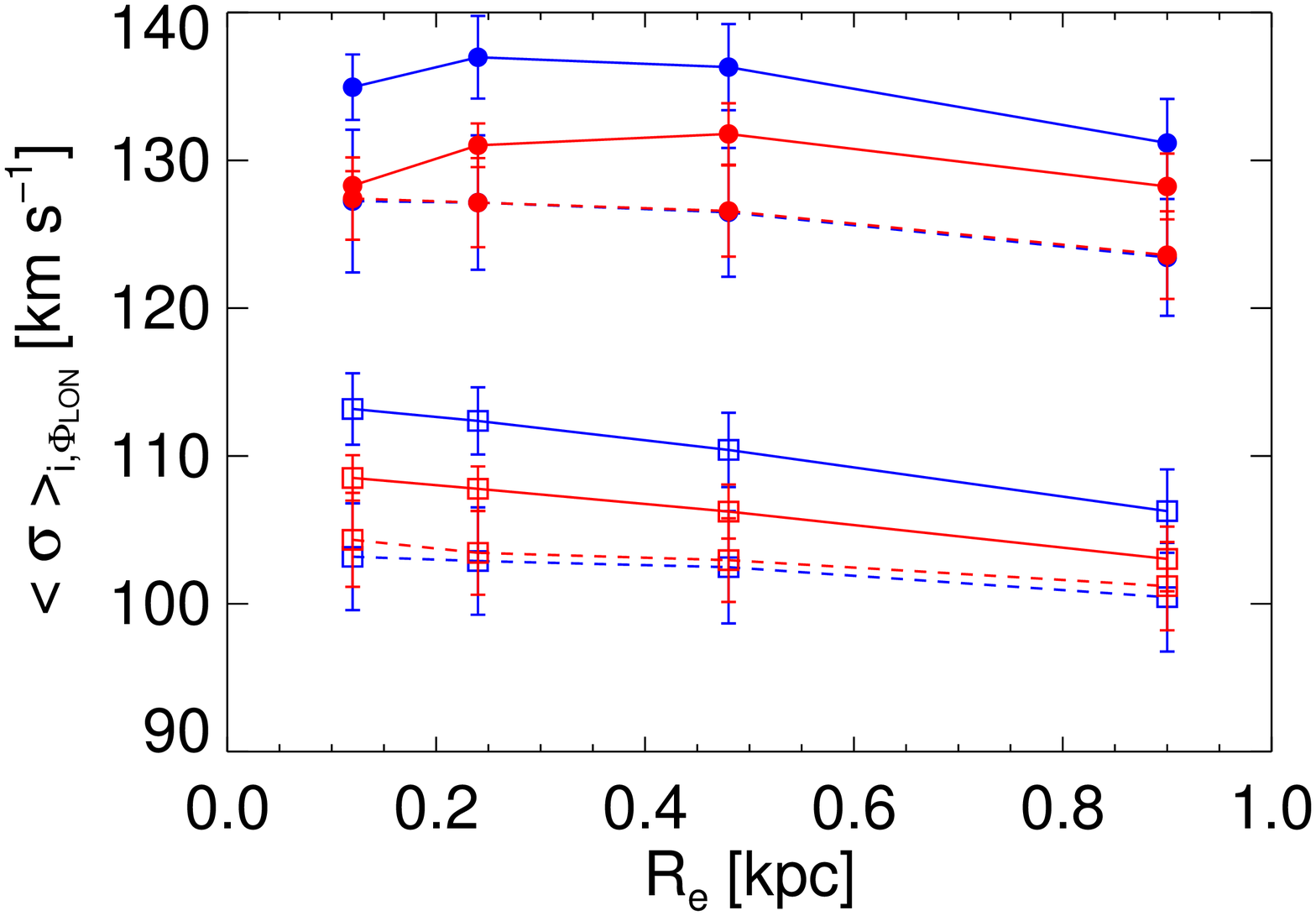}}

\centering{\includegraphics[scale=1.,width=.5\textwidth,trim=0.pt 0.pt 0.pt 0.pt,clip]{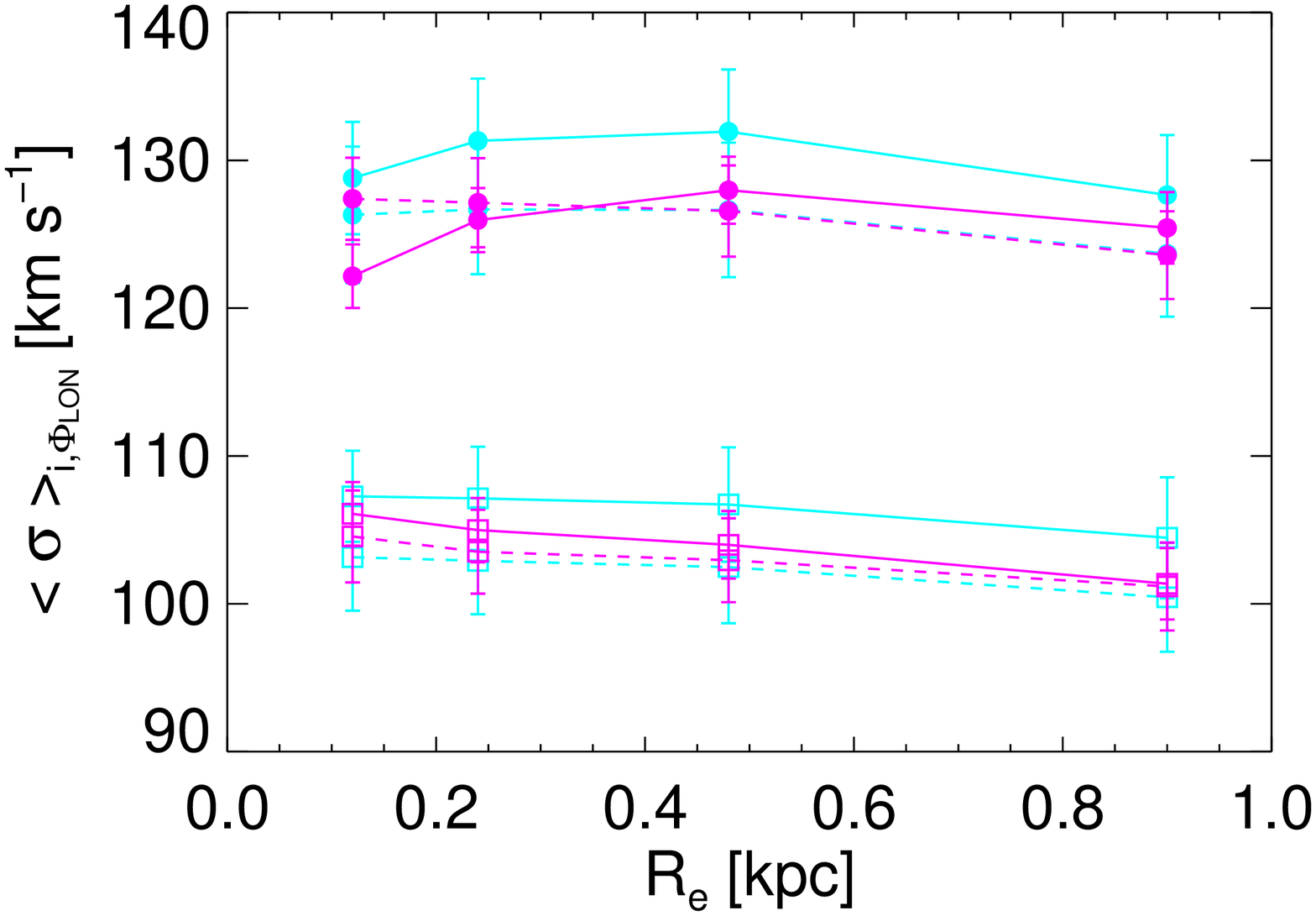}}

\caption{The orientation averaged RMS velocity $\sigoav$, measured for various assumed values of $\re= 0.12,0.24, 0.48, 0.9$~kpc.  Squares denote simulations with only a disk, while filled circles  denote disk+bulge simulations. Solid lines connect models with a black hole while dashed lines show models prior to the growth of a black hole; and blue/cyan curves and points denote barred models while red/pink denotes the unbarred models. The top panel shows results for  $\mcmc=10^8~\msol$, while the bottom panels show results for  $\mcmc =10^7~\msol$. For a given model (connected by lines), the value of $\sigoav$ is almost independent of $\re$ within $\re \sim 0.5$.} 

\label{fig:sig_re}

\end{figure}

Figure~\ref{fig:sig_re} shows how $\sigoav$ (the value of RMS velocity averaged over all orientations) varies with $\re$ for $\mcmc = 10^8\msol$ (top) and  $\mcmc = 10^7\msol$ (bottom).  $\sigoav$ depends slightly on  $\re$ in the disk-only simulation, but is almost independent of $\re$ for the disk+bulge model. Since $\sigoav$  is not strongly dependent on $\re$, hence hereafter we selected $\re=0.9$~kpc unless otherwise noted.  In the disk-only simulations this slightly overestimates the effective $R_e$ but the difference between the barred and unbarred systems is unlikely to be affected. The error bars represent the standard deviation obtained averaging over 9 different orientations. We emphasize that the error bars do not represent the error on the mean $\sigma$, but are meant to show the scatter introduced by orientation effects. While the error bars for the barred models with CMCs (blue/cyan solid curves) slightly overlap the error bars for the unbarred models (red/pink solid curves) it is clear that the mean values of $\sigoav$ for the barred models with CMCs are almost always larger by at least one standard deviation. This figure also shows that in the absence of the CMCs (dashed lines) there is little or no difference between the barred and unbarred galaxies, demonstrating that the orientation of the bar alone cannot be responsible for the observed differences.

We will discuss the vertical offsets between the different curves (corresponding to models with/without a bar, bulge, CMC in future sections).

\section{RESULTS}

\subsection{Factors Affecting the Measurement of $\sige$}
In this section we examine various factors that affect the observed $\sige$ in our simulations. These include the angle of the bar to the line-of-nodes (\S~\ref{sec:sige_philon}), the inclination of the disk to the line-of-sight (\S~\ref{sec:sige_inc}), and the growth and final mass of a CMC (\S~\ref{sec:sige_smbh}).

\subsubsection{Dependence of $\sige$ on $\philon$}
\label{sec:sige_philon}

\begin{figure*}
\centering{
\includegraphics[scale=1.,width=.45\textwidth,trim=0.pt 0.pt 0.pt 0.pt ,clip]{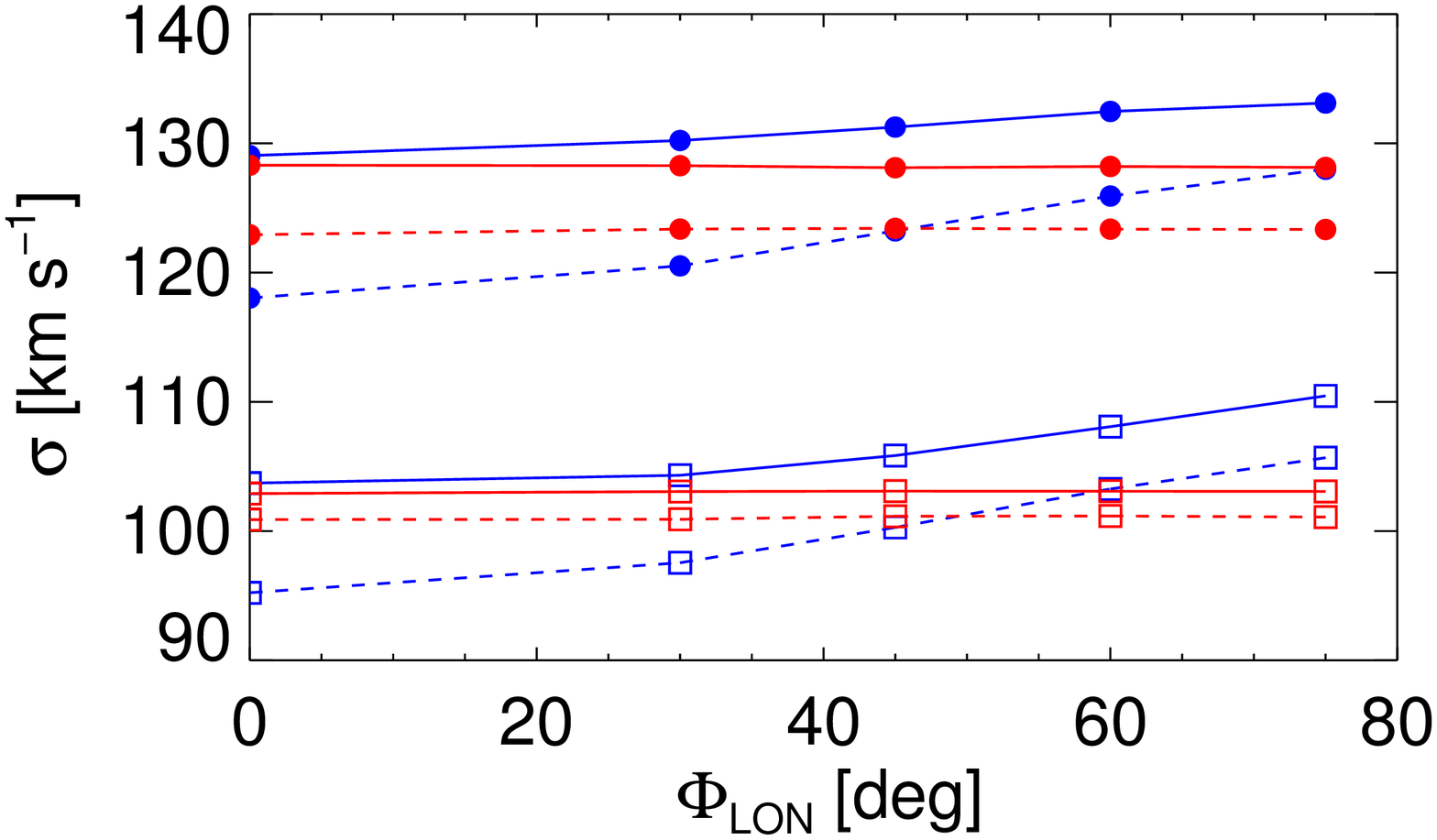}
\includegraphics[scale=1.,width=.45\textwidth,trim=0.pt 0.pt 0.pt 0.pt ,clip]{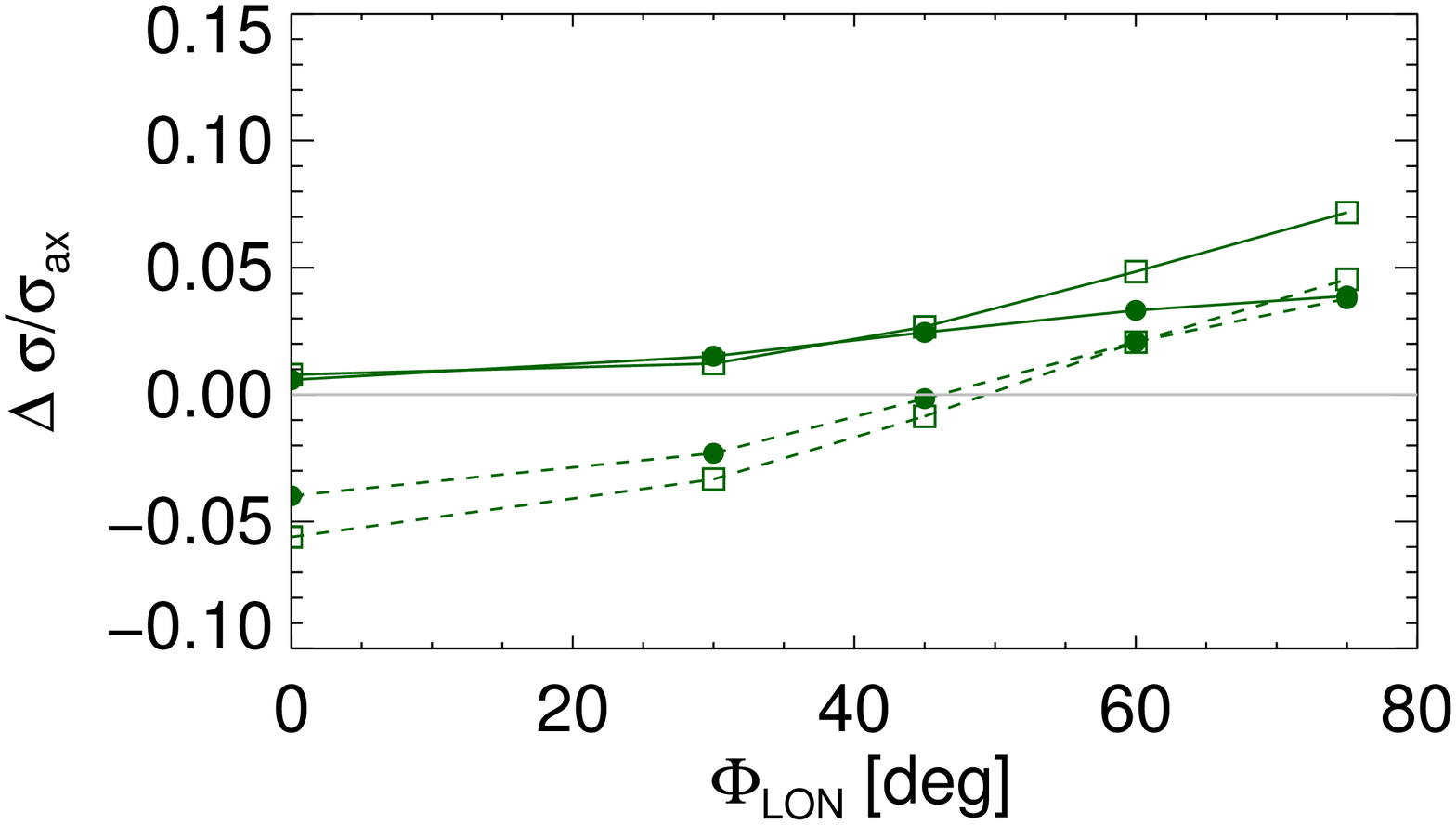}}

\centering{
\includegraphics[scale=1.,width=.45\textwidth,trim=0.pt 0.pt 0.pt 0.pt ,clip]{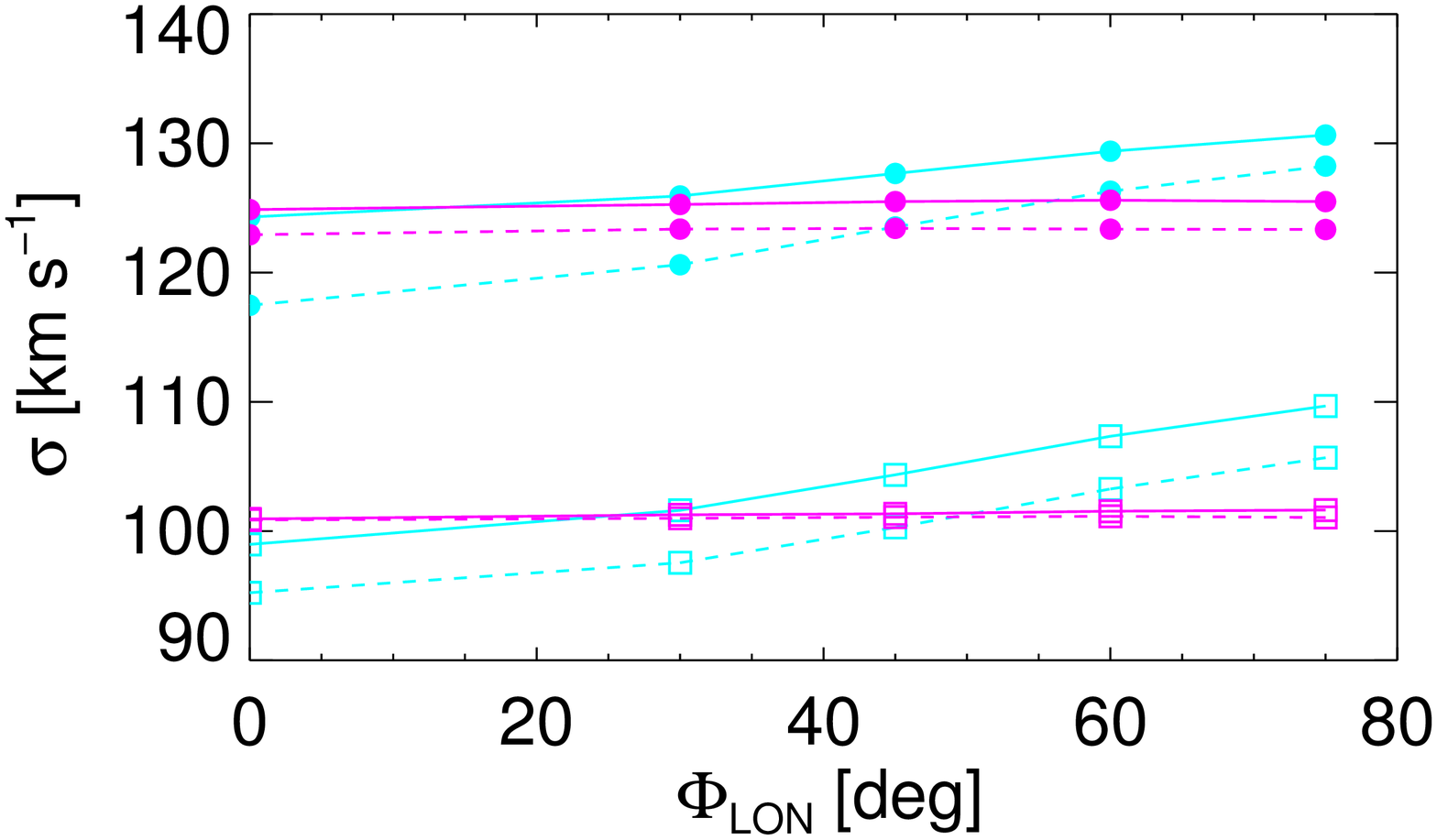}
\includegraphics[scale=1.,width=.45\textwidth,trim=0.pt 0.pt 0.pt 0.pt ,clip]{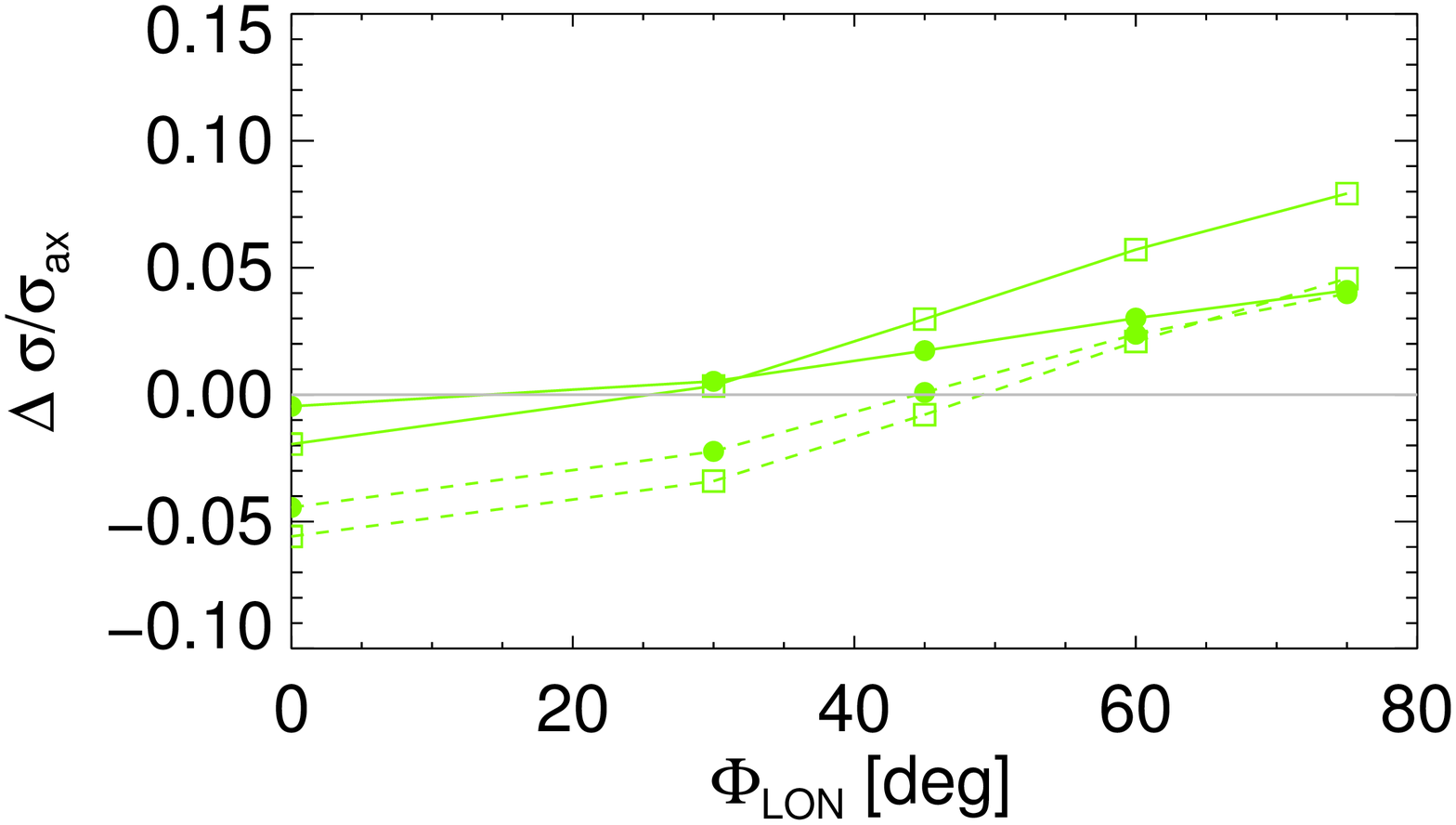}}

\caption{Left:  $\sige$ for $\re=0.9$~kpc, versus $\philon$, for each of our 8 datasets. The angle of inclination is fixed at $45^\circ$. Squares (circles) denote disk (disk+bulge) simulations, solid (dotted) lines denote the presence (absence) of a black hole, and blue/cyan (red/pink) denotes the presence (absence) of a bar. The top panels show the results for the $10^8~\msol$ CMC, while the bottom panel shows the results for the $10^7~\msol$ CMC. As expected, the unbarred (red) models show no dependence on $\philon$. The correlation between $\sige$ and $\philon$ in the barred cases is due to the alignment of the bar with our line of sight as $\philon$ approaches $90^\circ$. Right: The fractional change in velocity dispersion $\Delta\sige/\sigma_{\mathrm{ax}}$ (see text for definition) and different values of $\philon$ for $i=45^\circ$.  $\Delta\sige/\sigma_{\mathrm{ax}}$ increases as the bar is viewed more end-on.}
\label{fig:sig_philon}
\end{figure*}

\begin{figure*}
\centering{
\includegraphics[scale=1.,width=.45\textwidth,trim=0.pt 0.pt 0.pt 0.pt ,clip]{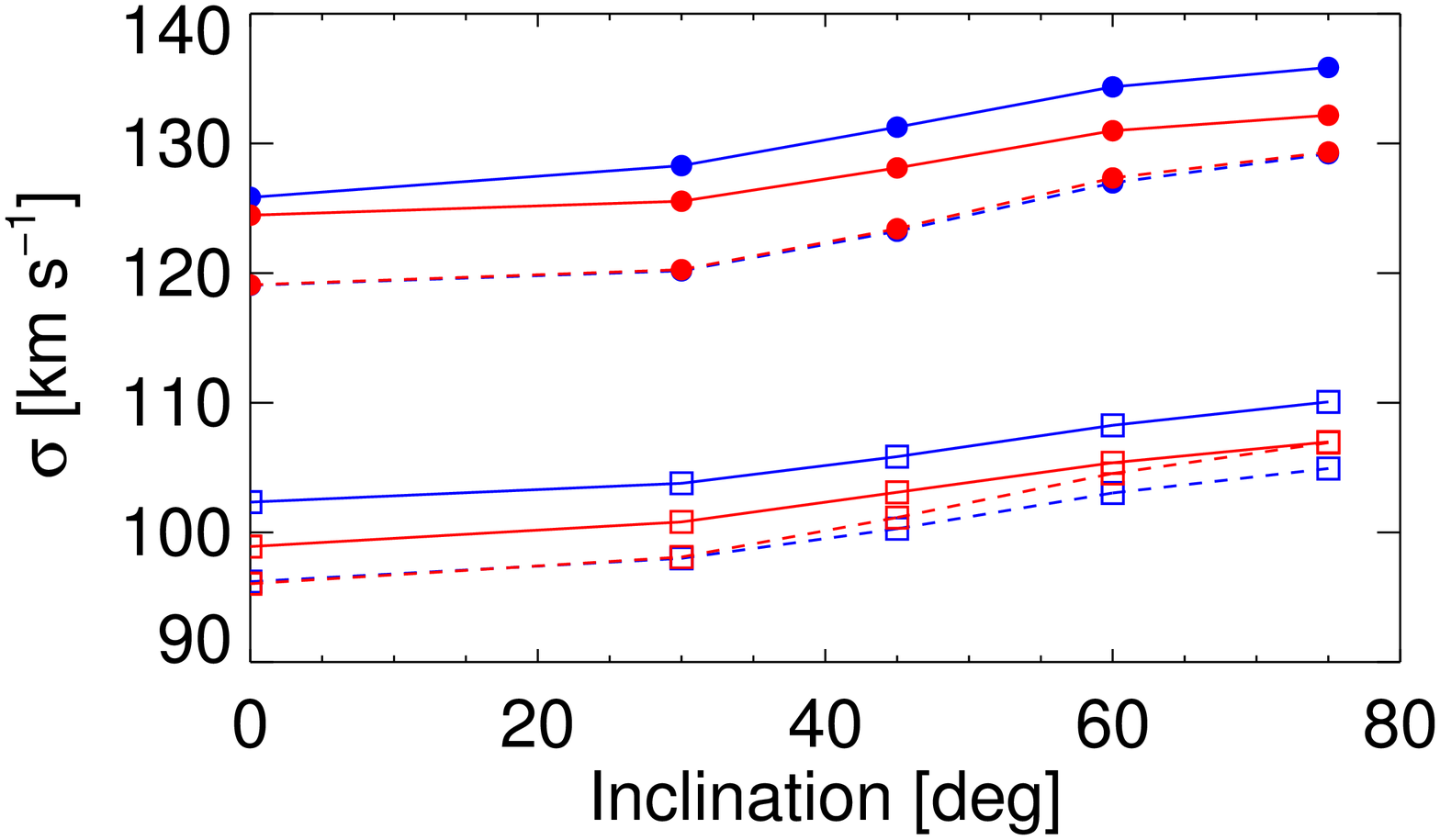}
\includegraphics[scale=1.,width=.45\textwidth,trim=0.pt 0.pt 0.pt 0.pt ,clip]{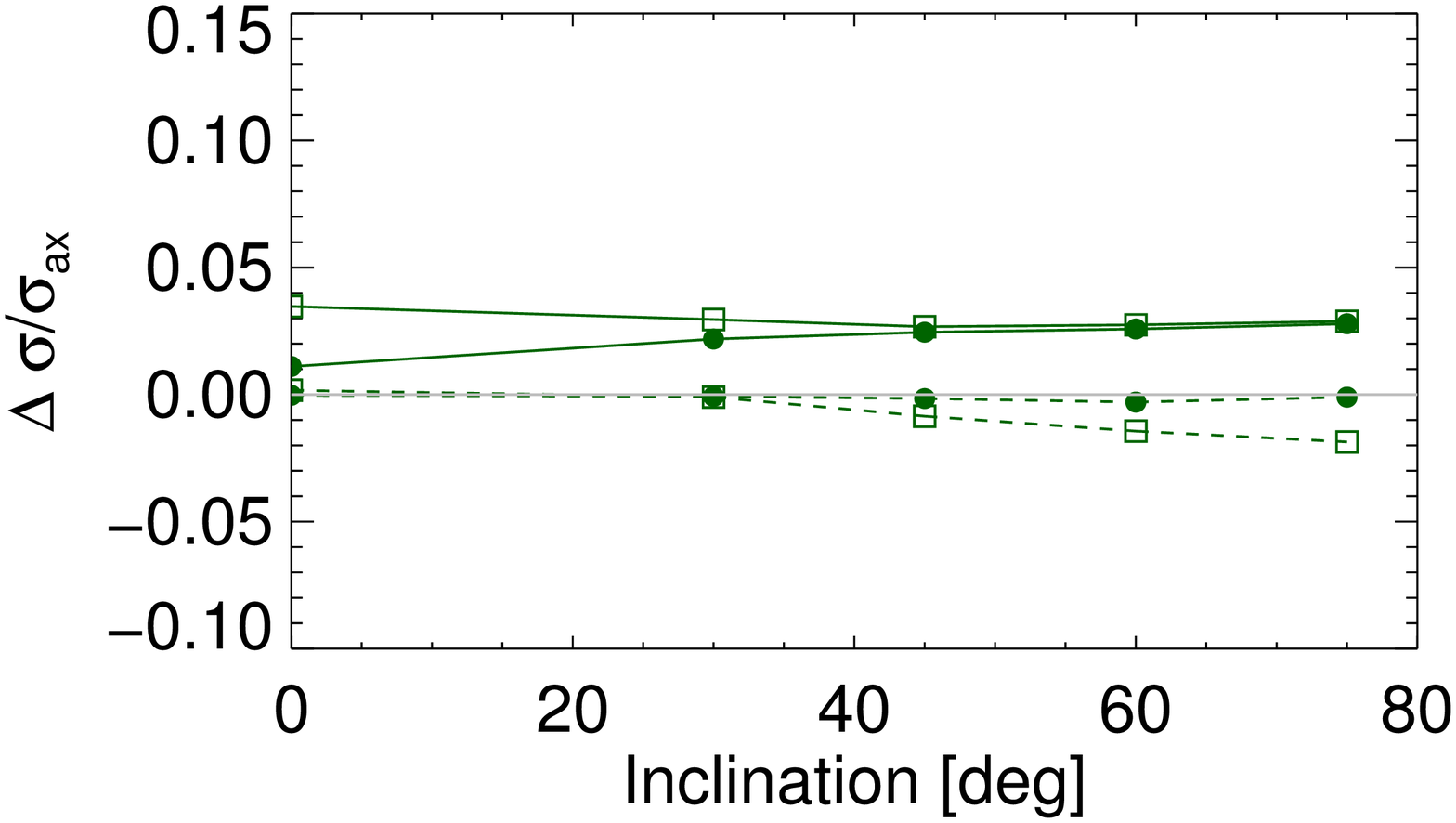}}

\centering{
\includegraphics[scale=1.,width=.45\textwidth,trim=0.pt 0.pt 0.pt 0.pt ,clip]{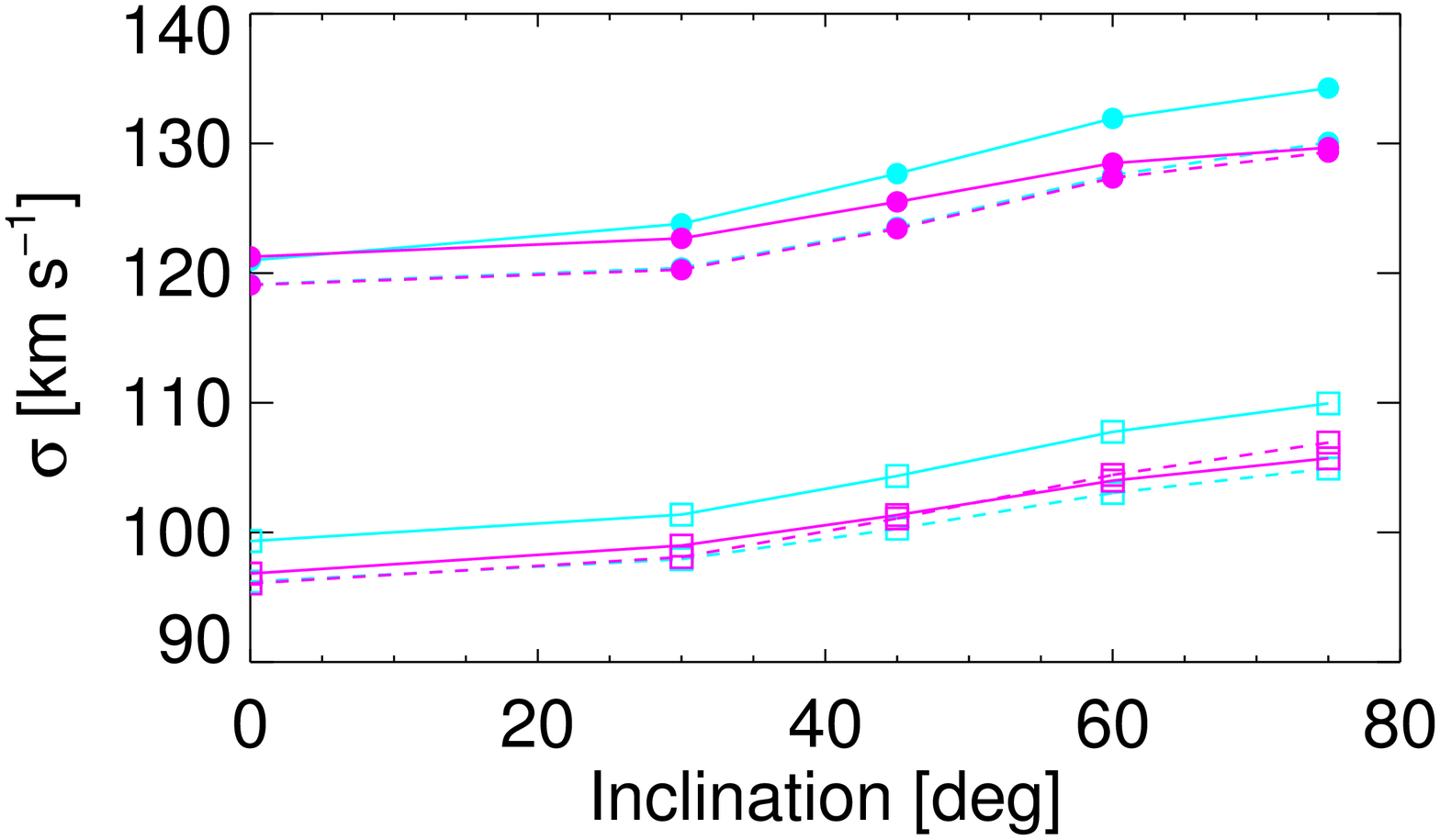}
\includegraphics[scale=1.,width=.45\textwidth,trim=0.pt 0.pt 0.pt 0.pt ,clip]{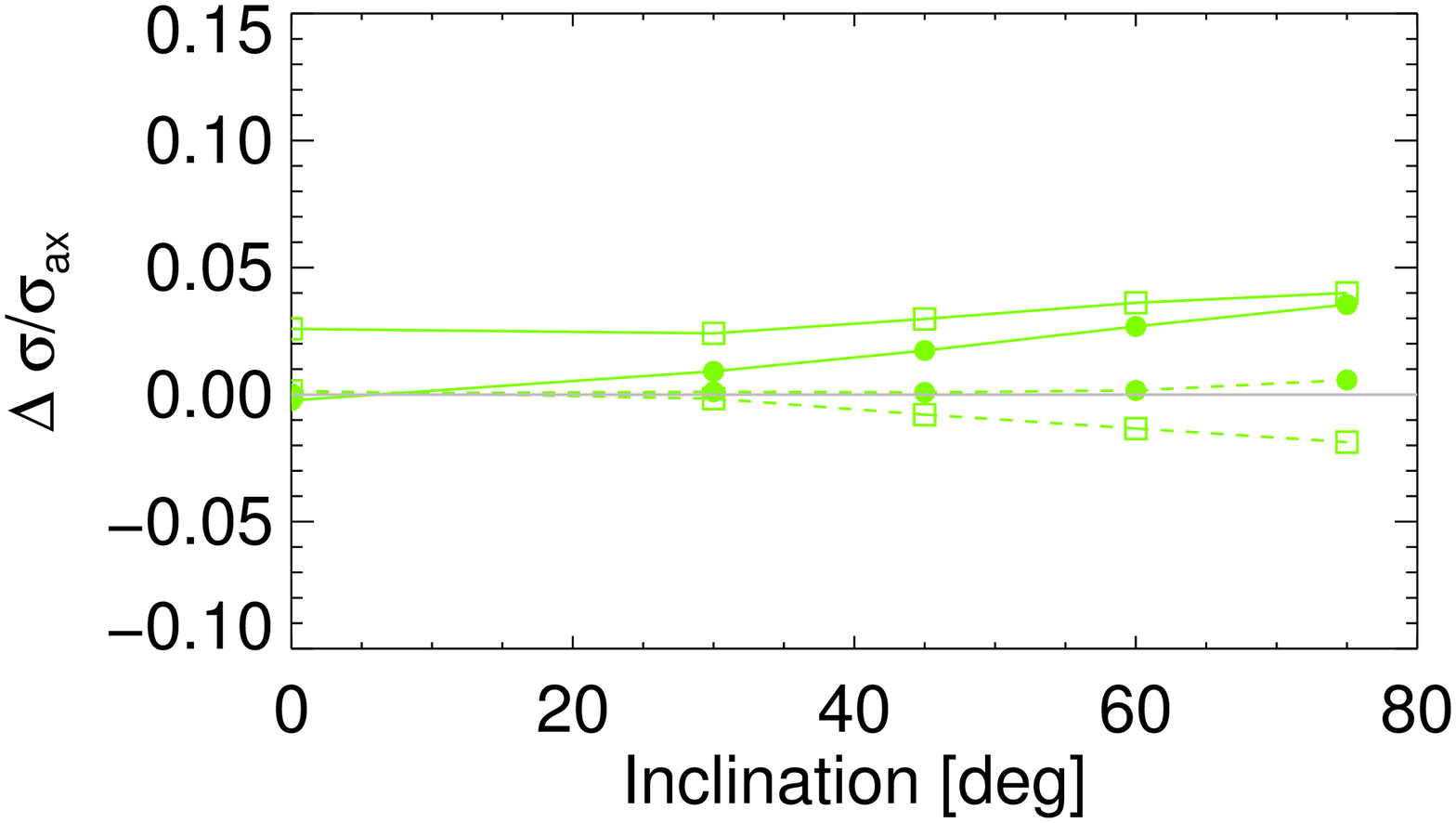}}

\caption{Left: $\sige$ for $\re=0.9$, versus the angle of inclination of the disk to the line of sight, $i$, for each of our 8 snapshots.  $\philon$ is fixed at $45^\circ$. The top panels show the results for $\mcmc = 10^8~\msol$, while the bottom panel shows the results for  $\mcmc = 10^7~\msol$. The measured $\sige$ increases with increasing inclination due to an increasing fraction of disk particles with anisotropic velocity dispersion contaminating the measurement of $\sige$. Right: The fractional change in velocity dispersion $\Delta\sige/\sigma_{\mathrm{ax}}$ as a function of inclination (see text for definition) for $\philon=45^\circ$. $\Delta\sige/\sigma_{\mathrm{ax}}$  is only weakly dependent on inclination.}
\label{fig:sig_inc}
\end{figure*}

Figure~\ref{fig:sig_philon}~(left) shows the dependence of $\sigreav$ on the choice of $\philon$, where $\sige$ is measured within $\re = 0.9$~kpc. The angle of inclination of the disk is fixed at $45^\circ$. In the barred cases, the positive correlation between $\sige$ and $\philon$ is to be expected from a simple geometrical argument. Bar supporting $x_1$ orbits are elongated along the bar, and their primary motion is oscillation back and forth along its major axis \citep[e.g.][]{Sellwood93,Athanassoula92,Bureau99,Shen04}. 
In the disk-only cases (blue/cyan squares) we see that as the orientation of the bar approaches end-on (i.e. as $\philon \rightarrow 90^\circ$ and the major axis of the bar aligns with the line-of-sight) $\sige$ increases. This is because a given circular aperture of radius $\re$ encloses a greater fraction of $x_1$ orbits for end-on bar orientations. The alignment of these radial orbits with the line-of-sight results in a wider distribution of line-of-sight velocities, increasing our measurement of $\sige$. \citet{Shen04} showed for similar disk-only simulations that the $x_1$ family which supports the bar is slowly destroyed by a growing CMC. However, they found that the mass of the CMC necessary to completely destroy this family (and the bar) was about 25 times larger than the most massive CMC used in our simulations. 

In the unbarred counterpart (red/pink squares) all the disk particles have been scrambled in azimuth as described in \S~\ref{sec:mod_set}, erasing the bar, but preserving the radially averaged mass and kinematic profiles. For the unbarred models $\philon$ is not defined (since there is no bar with respect to which the angle of the line-of-nodes can be measured), however to make it clear that the velocity dispersion is constant for all line-of-sights with the same inclination, we  mark the measured $\sige$ by red/pink squares or solid dots connected by horizontal lines. At time $t_2$ following the growth of the CMC, the initially unbarred models develop weak spiral patterns which cause small dependence on $\philon$ which we show connected by solid red/pink lines.

Before the CMC is grown, the barred simulation with the classical bulge (solid blue{/cyan} dots connected by dashed curves) shows a dependence on $\philon$ similar to the disk-only case  (open blue{/cyan} squares connected with dashed curves). The vertical offset of the former results because of the added mass of the bulge. However, after the growth of the CMC (solid blue{/cyan} dots and lines) the dependence on $\philon$ is significantly weaker in the presence of the bulge than in the absence of the bulge. This implies that when the CMC grows inside a bulge+bar it results in a more significant reduction in the fraction of $x_1$ orbits, compared to when the identical CMC grows in a pure bar.  We investigate the cause of this in Section~\ref{sec:ang_mom}.

Figure~\ref{fig:sig_philon}~(right) shows the fractional difference $\Delta\sige/\sige_{\mathrm{ax}} = (\sige_{\mathrm{bar}} - \sige_{\mathrm{ax}})/ \sige_{\mathrm{ax}}$ between $\sige$ for a barred model and its unbarred counterpart, relative to the unbarred case. $\Delta\sige/\sige_{\mathrm{ax}}$ is plotted as a function of $\philon$ (while keeping the inclination fixed at $i=45^\circ$).  For the models without a CMC (dashed lines) the orientation of the bar can result in either {\em negative} $\Delta\sige/\sige_{\mathrm{ax}}$ as the bar becomes parallel to the line-of-nodes or {\em positive}  $\Delta\sige/\sige_{\mathrm{ax}}$ as the bar is viewed end-on. In contrast, after the growth of a CMC (solid curves) $\sige$ is always larger for the barred case than for the unbarred case regardless of the value of $\philon$, but once again the fractional difference becomes larger as the bar is seen end-on (i.e. $\philon \rightarrow 90^\circ$). In the presence of a classical bulge (solid-dots) the maximum difference in $\sige$ is about 5\%. Interestingly, the fractional difference in $\sige$ is larger in the {\it absence} of a classical bulge, but even so it is $\lesssim 10 \%$. A comparison of the top panel ($\mcmc = 10^8\msol$) and bottom panel ($\mcmc = 10^7\msol$) shows that the overall trends are similar for the 2 CMCs. In fact the right hand panels in this figure show that there is almost no difference in $\Delta\sige/\sige_{\mathrm{ax}}$ despite the fact that the CMC in the bottom-right panel is a factor of 10 smaller than that in the top-right panel.

\subsubsection{Dependence of $\sige$ on Inclination }
\label{sec:sige_inc}

Figure~\ref{fig:sig_inc} shows the dependence of $\sigreav$ on the angle of inclination of the disk to the line-of-sight ($\philon$= $45^\circ$). Once again the dependence of $\sige$ on inclination can, in part, be explained with a geometrical argument. At low inclination (i.e. nearly face-on) the contribution of the rotational velocity component of the disk to $\sige$ is relatively insignificant.  However, the number of disk particles contained within a given aperture of radius $\re$ increases with inclination. As the inclination increases, a larger number of disk particles on both the near and far side of the nuclear region fall within $\re$, causing  $\sige$ to increase. Note that if the disk orbits were perfectly circular the orbits falling within $\re$ would have velocities which are nearly perpendicular to the line-of-sight and would have little effect on $\sige$. However since the orbits in the inner region of both the barred and scrambled disks are quite radial, there is a fairly strong dependence on inclination, for both the barred (blue/cyan) and scrambled (red/pink) models (see Fig.~\ref{fig:sig_inc}~left).

There is also a more subtle contribution to the correlation between $\sigreav$ and inclination. As inclination increases and the orientation of the disk becomes more edge-on, the intrinsic (3-dimensional) velocity dispersion becomes dominated by the radial and tangential dispersions, $\sigr$ and $\sigp$ respectively, rather than the vertical dispersion $\sigz$. As we will show in Figure~\ref{fig:sig_beta},  $\sigr$ and $\sigp$ are greater than $\sigz$, contributing to a positive correlation between $\sigreav$ and inclination. 

Figure~\ref{fig:sig_inc}~(right) shows the fractional difference $\Delta\sige/\sige_{\mathrm{ax}}$ as a function of inclination (with  $\philon=45^\circ$).  For the models without a CMC (dashed lines) $\Delta\sige/\sige_{\mathrm{ax}}$  is almost independent of inclination. After the growth of a CMC (solid curves) $\sige$ is larger for the barred cases than for the unbarred cases  (i.e. both solid curves are above zero for all values of $i$) and depends weakly on inclination. In the presence of a classical bulge (solid-dots connected by solid lines) the maximum increase in $\sige$ is about 3\% for a nearly edge-on orientation, and about 4\% for the pure disk (squares connected by solid lines).

It is important to note that $\philon$ is fixed at $45^\circ$, hence the orientation of the bar can essentially be thought of as intermediate between the side-on and end-on orientations. It was evident in Figure~\ref{fig:sig_philon} that a side-on view of the bar produces values of $\sigreav$ which are less than the unbarred case, while an end-on view of the bar does the opposite. Thus, when $\philon$ is fixed at $45^\circ$, the barred and unbarred observations at $t_1$ produce nearly identical values of $\sigreav$. This allows for a direct comparison between the $t_2$ values of $\sigreav$ in the barred and unbarred cases. The growth of a CMC in the presence of a bar clearly produces a greater change in $\sigreav$ than the growth of the same CMC in an unbarred galaxy.

\subsubsection{Dependence of $\sige$ on CMC Growth}
\label{sec:sige_smbh}

In Section~\ref{sec:sige_philon} and  Section~\ref{sec:sige_inc} we saw that for both the unbarred and barred models $\sige$ is more sensitive to the presence/absence of a CMC than to changes in the orientation of the disk to the line of sight. This is  surprising since the sphere of influence of the CMC (estimated to be $\sim 0.17$~kpc for models with $\mcmc = 10^8 \msol$) {\em is a factor of six smaller than $\re = 0.9$~kpc!} This implies that the gravitational potential of the CMC is not directly responsible for this increase, rather it is the effect that the changing potential has on the evolution of the bar. In this subsection we quantify the effect of the growth of the CMC on $\sige$ and in the following  two sections we examine the causes of this increase.

\begin{figure*}
\centering{
\includegraphics[scale=1.,width=.45\textwidth,trim=0.pt 0.pt 0.pt 0.pt ,clip]{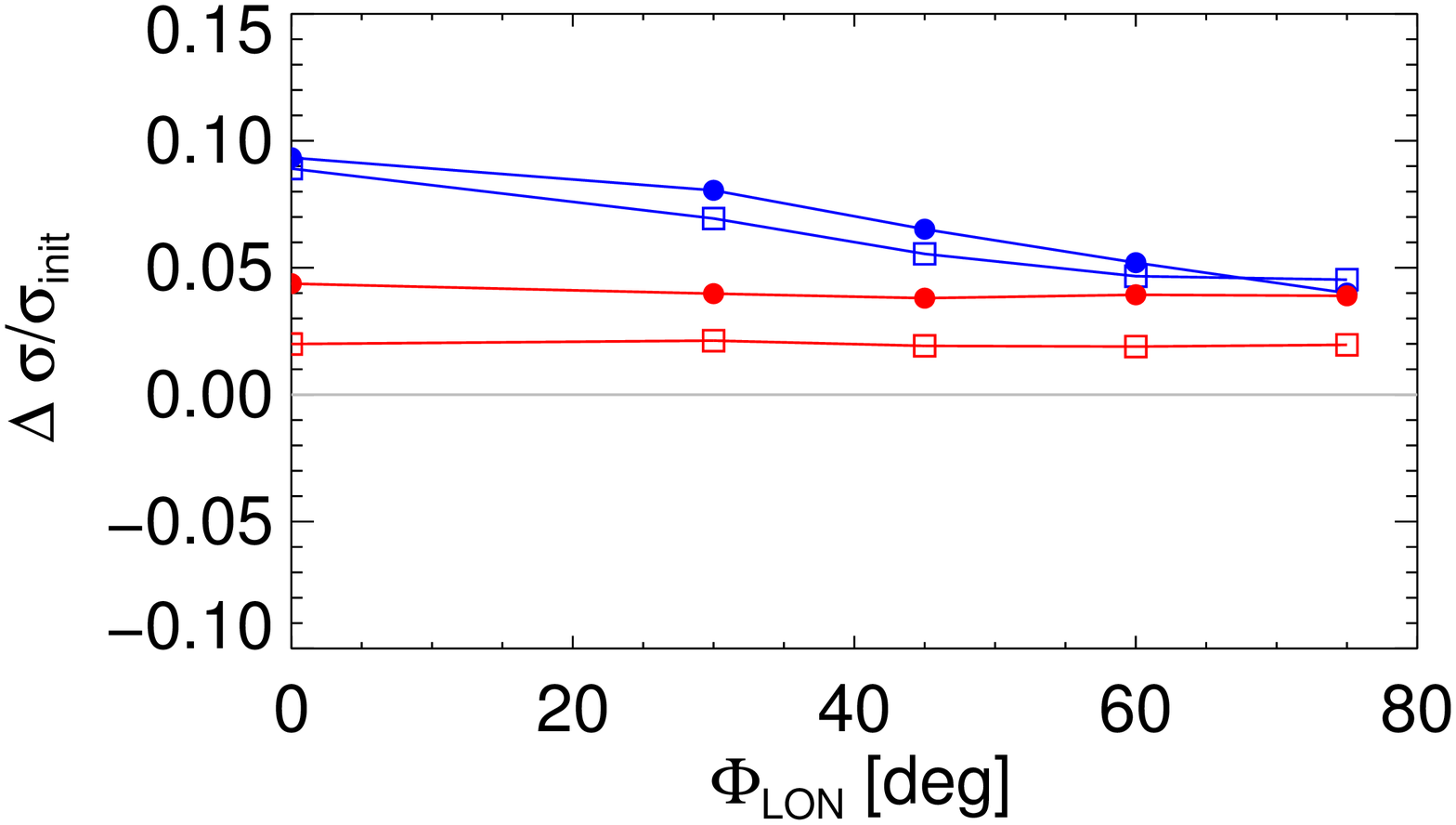}
\includegraphics[scale=1.,width=.45\textwidth,trim=0.pt 0.pt 0.pt 0.pt ,clip]{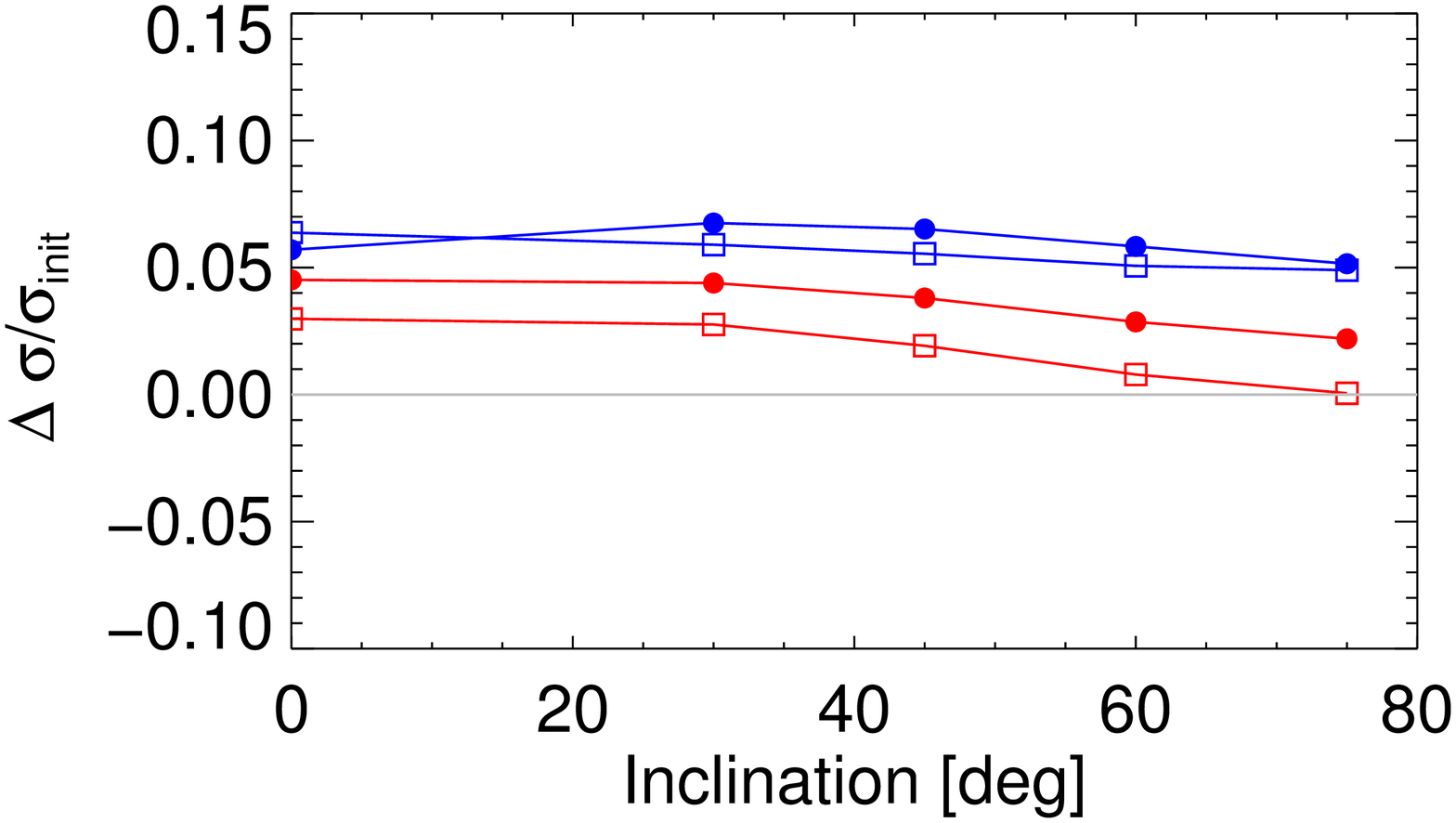}}

\centering{
\includegraphics[scale=1.,width=.45\textwidth,trim=0.pt 0.pt 0.pt 0.pt ,clip]{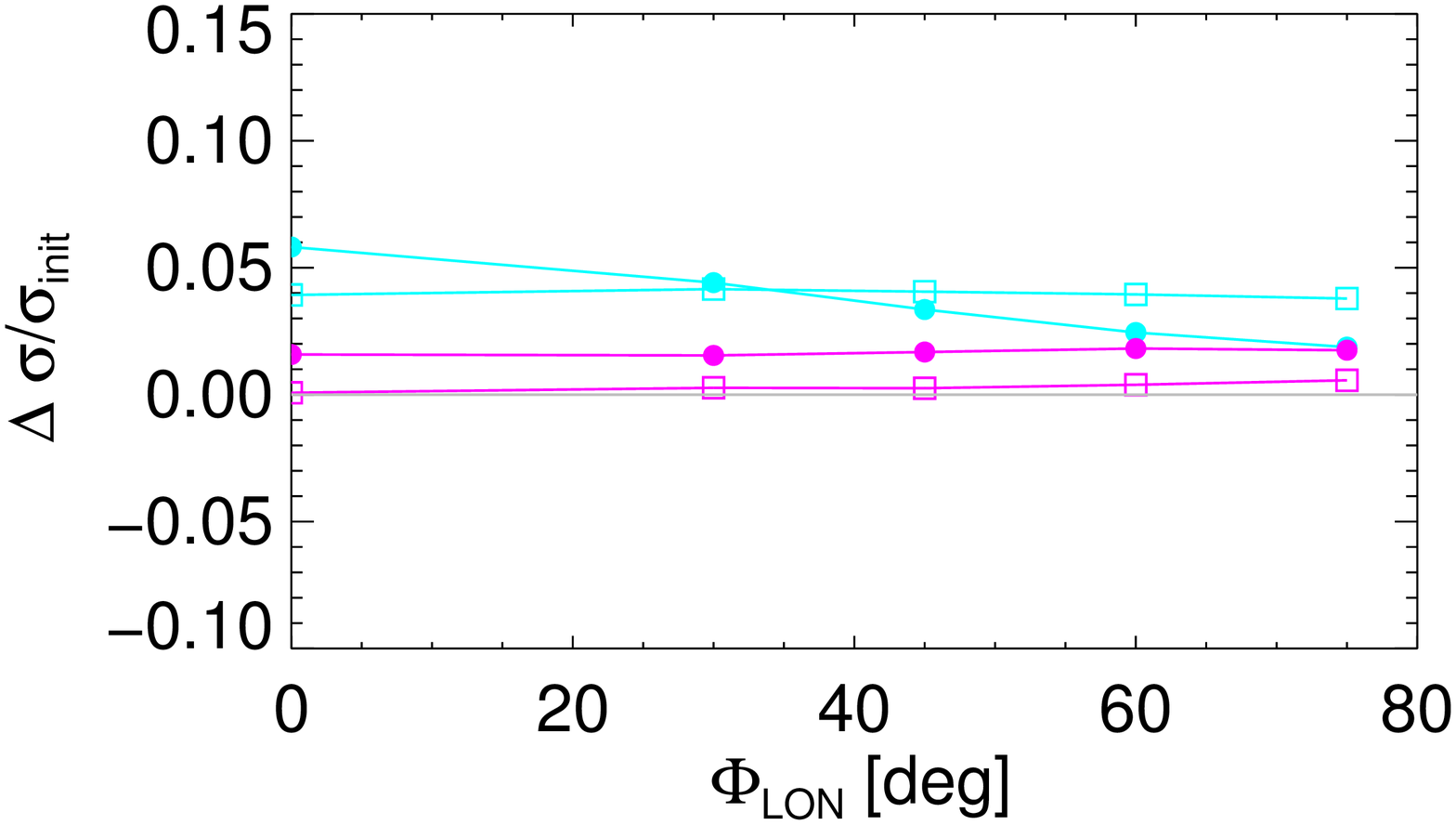}
\includegraphics[scale=1.,width=.45\textwidth,trim=0.pt 0.pt 0.pt 0.pt ,clip]{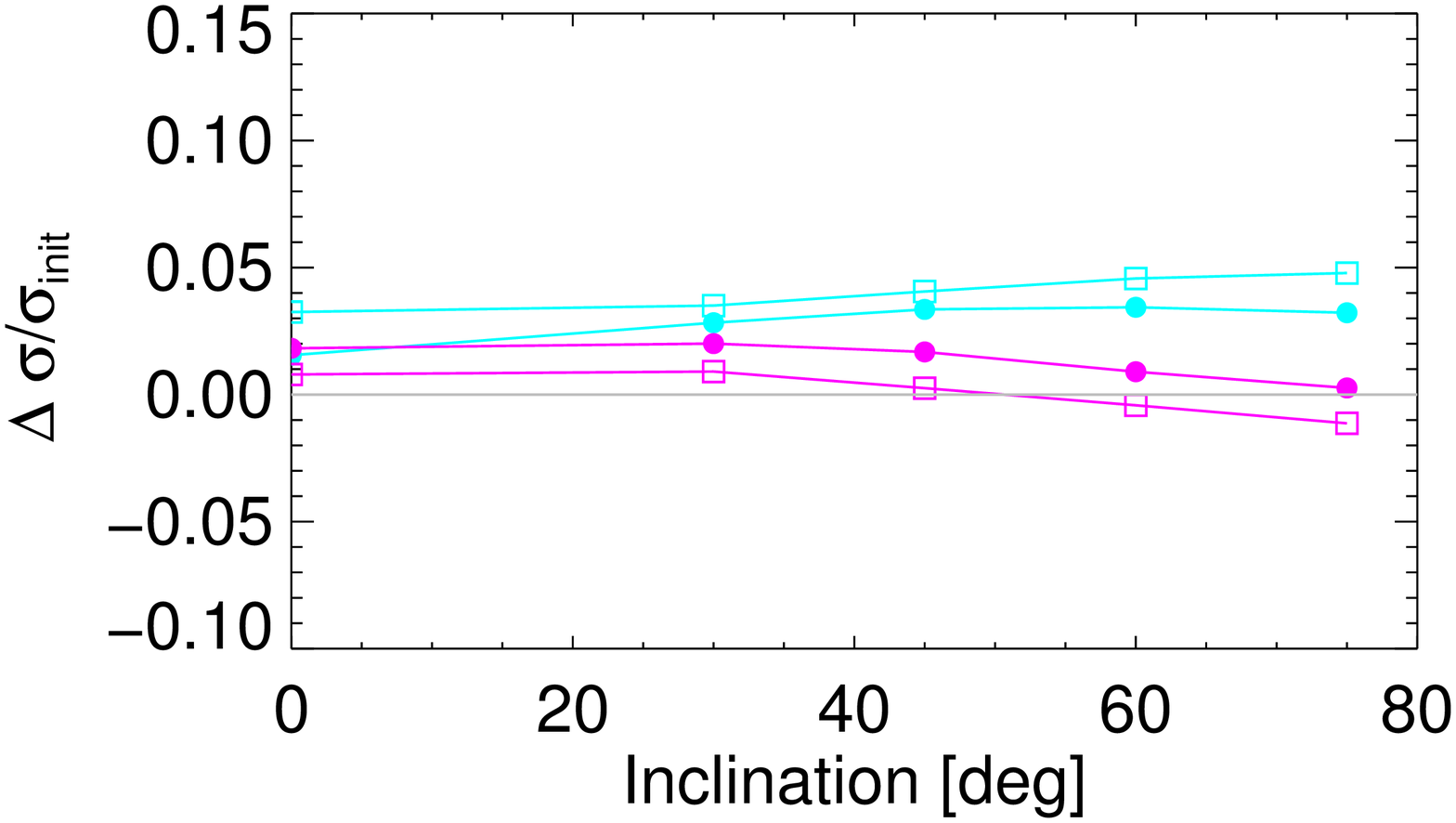}}

\caption{Fractional change in  velocity dispersion $\Delta \sige/\sige_{\mathrm{init}}$  (see text for definition) due to the growth of the CMC as a function of $\philon$ when inclination is fixed at $45^\circ$ (left), as a function of $i$ when $\philon=45^\circ$ (right). The top panels show results for $\mcmc=10^8~\msol$, while the bottom panel shows results for  $\mcmc=10^7~\msol$. The square (circle) symbols represent disk (disk+bulge) simulations. Blue/cyan lines indicate that the system is barred and red/pink lines are for the unbarred models.
}  
\label{fig:sige_smbh}
\end{figure*}

Figure~\ref{fig:sige_smbh} shows  the fractional change $\Delta \sige/\sige_{\mathrm{init}} = (\sige(t_2) -\sige(t_1))/\sige(t_1)$ ) for the unbarred (red/pink) and barred (blue/cyan) models without (squares) and with (solid dots) a classical bulge.  $\Delta \sige/\sige_{\mathrm{init}}$  is plotted as a function of $\philon$ for models with $i=45^\circ$  (Fig.~\ref{fig:sige_smbh}~left) and as a function of inclination for models with $\philon=45^\circ$ (Fig.~\ref{fig:sige_smbh}~right). In the left panel we see that in the unbarred models the growth of the CMC produced an increase in $\sige$ ($\sim 3-5$\% when $\mcmc = 10^8 \msol$) that  is essentially independent of $\philon$ (the very small fluctuations with $\philon$ arise from the weak spiral features in the unbarred models at $t_2$).  

The barred models (blue) display a larger relative increase in $\sige$ ($\sim 5-10$\% when $\mcmc = 10^8 \msol$). In Figure~\ref{fig:sig_philon}~(left) we saw that the growth of the CMC in a bulge+bar model (solid blue dots) results in no dependence on $\philon$. This implies that the velocity distribution of stars within $\re$ is essentially isotropic. It appears that the growth of a CMC scatters and therefore axisymmetrized a significant portion of bar supporting orbits in the inner most regions of the system.  This results in a reduced dependence of $\sige$ on $\philon$ in the barred disk+bulge simulations at $t_2$. Thus $\Delta \sige/\sige_{\mathrm{init}}$ decreases with increasing $\philon$. The weakening of the bar is less significant in the disk-only simulation, resulting in a flatter relationship between $\Delta \sige/\sige_{\mathrm{init}}$ and $\philon$.

In the right hand panels of Fig.~6  all models tend to show a similar dependence on inclination (with $\philon=45^\circ$). The increase in $\Delta \sige/\sige_{\mathrm{init}}$ following the growth of a CMC is inversely proportional to the inclination. This trend is evident in the unbarred simulations, and, to a lesser extent, in the barred disk+bulge simulation.  As inclination is increased,   the fractional change in $\sigma$ between $t_1$ and $t_2$ decreases. A comparison of top and bottom panels shows that the larger CMC (top)  produces a 2-3\% larger increase in $\Delta \sige/\sige_{\mathrm{init}}$ only for $\philon \sim i\sim 30^\circ$. For other orientations we see almost no dependence on the mass of the CMC.
 
We note that the axisymmetric disk-only simulation with the $10^7 \msol$ CMC shows a slight decrease in $\sige$ between $t_1$ and $t_2$ at high inclination. In Figure~\ref{fig:sig_re} this  simulation also showed a decrease in $\sige$ between $t_1$ and $t_2$ at small $\re$. Both of these trends can be attributed to a significant decrease in radial velocity dispersion after the CMC is grown. This decrease radial dispersion is most prominent at small radii and causes the decrease in $\sige$ in this simulation for small values of $\re$. We therefore conclude that a less massive CMC mass will produce a slightly smaller increase in $\sige$ than a more massive CMC, but will nonetheless produce an increase that is larger in a barred disk than in an unbarred disk.

We defer a discussion of the cause of the decrease in $\Delta \sige/\sige_{\mathrm{init}}$ with increasing inclination to the next section (see Figure~\ref{fig:sig_beta}), where we show that this is because the intrinsic velocity dispersions in the radial, azimuthal and vertical directions ($\sigr$, $\sigp$, and $\sigz$) all increase by roughly the same amount.

\subsection{Velocity Dispersion and Velocity Anisotropy Profiles}

\begin{figure*}
\centering{\includegraphics[scale=1.,width=\textwidth,trim=0.pt 225.pt 0.pt 125.pt ,clip]{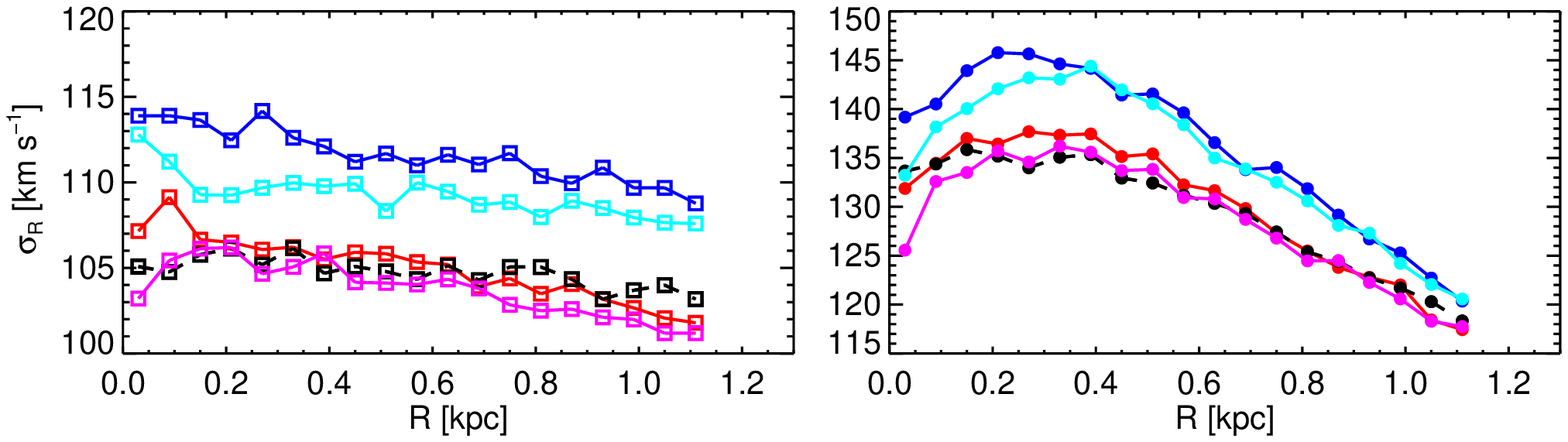}}
\centering{\includegraphics[scale=1.,width=\textwidth,trim=0.pt 225.pt 0.pt 125.pt ,clip]{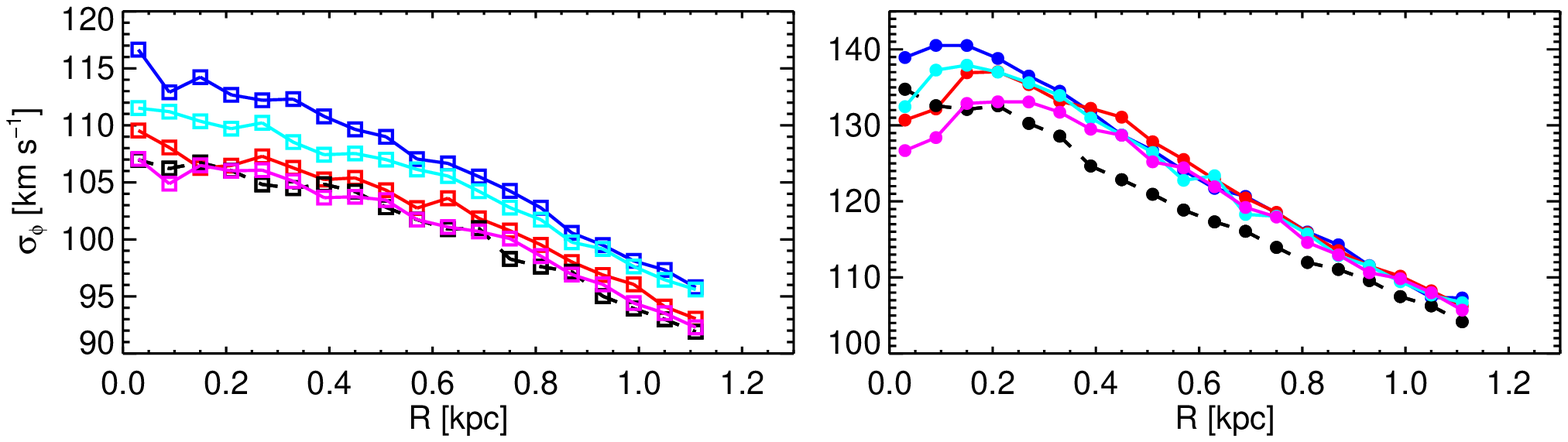}}
\centering{\includegraphics[scale=1.,width=\textwidth,trim=0.pt 225.pt 0.pt 125.pt ,clip]{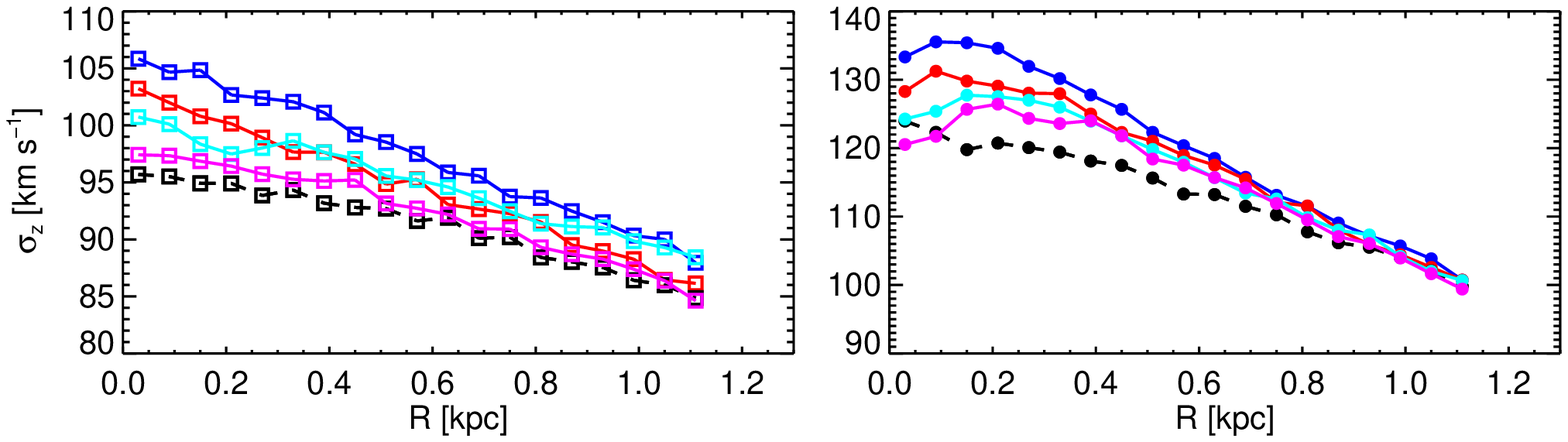}}
\centering{\includegraphics[scale=1.,width=\textwidth,trim=0.pt 225.pt 0.pt 125.pt ,clip]{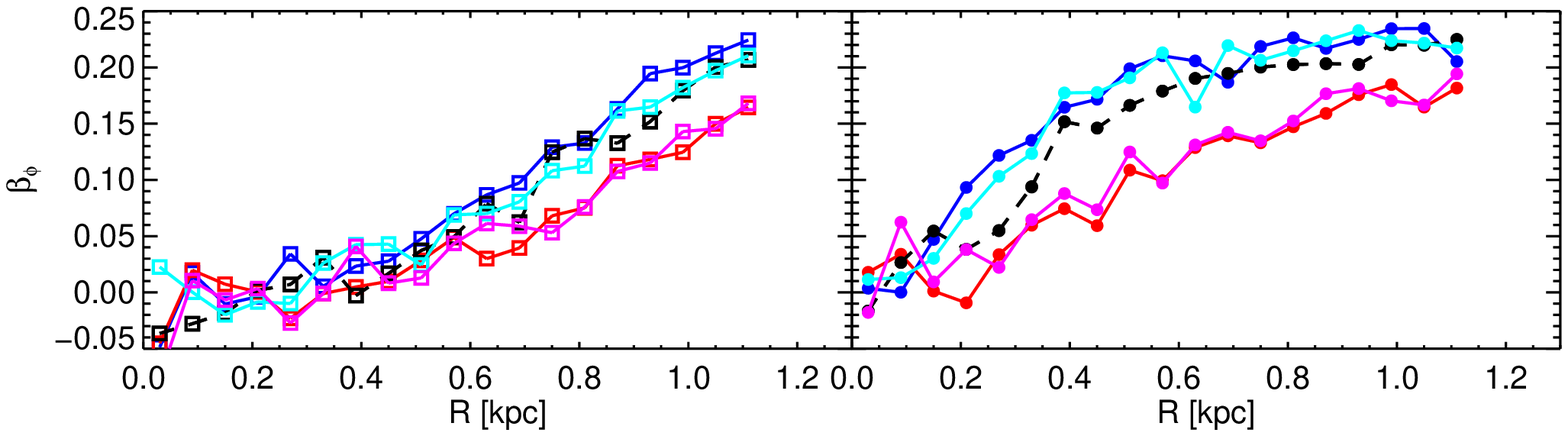}}
\centering{\includegraphics[scale=1.,width=\textwidth,trim=0.pt 225.pt 0.pt 125.pt ,clip]{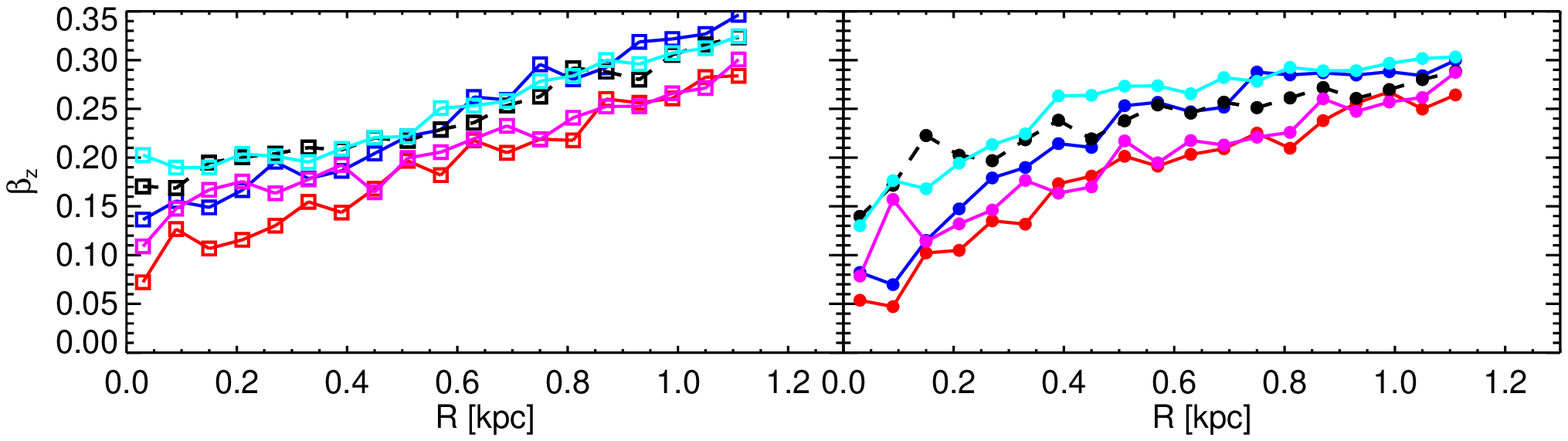}}
\caption{As a function of radius from top to bottom:  $\sigr$, $\sigp$,  $\sigz$, $\betap$ and $\betaz$. disk-only simulations are shown in the left panel; disk+bulge simulations are shown in the right panel. Solid (dotted) lines denote the presence (absence) of a black hole, and blue/cyan (red/pink) denotes the presence (absence) of a bar, while black curves/points show the velocity distributions of both barred and unbarred models at $t_1$.}
\label{fig:sig_beta}
\end{figure*}

To analyze the distributions of intrinsic velocity anisotropy, we compute the standard deviation of the radial, tangential, and vertical particle velocity distributions $\sigr$, $\sigp$, and $\sigz$ of particles enclosed within cylindrical annular bins in $R$. The bins have a width of 0.06~kpc and contain $\gtrsim 10^4$ particles on average.  We use these quantities to compute  the tangential anisotropy parameter $\betap = 1 - \sigp^2/\sigr^2$ and vertical anisotropy parameter $\betaz = 1 - \sigz^2/\sigr^2$ as a function of radius.  Figure~\ref{fig:sig_beta} shows (from top to bottom) $\sigr$, $\sigp$,  $\sigz$, $\betap$ and $\betaz$ as a function of cylindrical radius $R$.  These quantities are shown for the disk-only models (left), and disk+bulge models (right). Recall that anisotropy values $\betap=0, \betaz=0$ signify that $\sigp=\sigr$ and $\sigz = \sigr$ respectively. A positive value of $\beta$ signifies a larger radial velocity dispersion. 

In all cases at time $t_1$, the barred and unbarred models overlap due to the fact that their cylindrically averaged velocity ellipsoids are identical, and are therefore represented by the dotted black lines. While we recognize that cylindrically averaging the barred models erases physically important non-axisymmetric features in the shapes of the velocity ellipsoids, we are justified in doing this because the main purpose of these figures is to understand the differences in the measured values of $\sige$  which themselves are obtained by averaging over a circular region of projected radius $\re$.

The bottom two rows of Figure~\ref{fig:sig_beta} show that in both the disk-only (left) and disk+bulge (right) simulations, the growth of a CMC in an unbarred potential  (red curves)  definitively reduces both $\betap$ and $\betaz$ relative to the models at $t_1$ (black curves) over most of the radial range plotted. From examining the top three rows is clear that the decreases in $\betap$ and $\betaz$ are because  $\sigp$ and $\sigz$ increase slightly between $t_1$ and $t_2$, but $\sigr$ (top row) remains essentially unchanged, or even decreases slightly between  $t_1$ and $t_2$  This comes as no surprise given previous studies \citep[e.g.][]{Goodman84, quinlan_etal_95, Sigurdsson04} which show that the adiabatic growth of a CMC in  an axisymmetric system preferentially increases $\sigp$ over $\sigr$, thus reducing radial anisotropy. We see here that $\sigz$ also increases quite significantly relative to $\sigr$, resulting in a decrease in $\betaz$. Notice that the increase in $\sigr$ and $\sigp$ due to the growth of the a CMC with $\mcmc = 10^8\msol$ in the barred galaxies (blue) are slightly larger than the increase due to the smaller CMC (cyan). But in the unbarred simulations (red/pink) the difference resulting from the two CMCs is negligible and both are similar to the initial values of $\sigr$ and $\sigp$ (black curves). For the unbarred galaxies only $\sigz$ differs from the initial models.

The growth of both CMCs in the barred simulations results in a significantly larger increase in the radial velocity dispersion than in the corresponding unbarred cases. This is seen in the top row of Figure~\ref{fig:sig_beta}, which show the blue/cyan curves in both the disk-only (left) and disk+bulge (right) models to be significantly higher than for the initial models at $t_1$ (black curves) and the unbarred models after the growth of the CMC (red/pink curves). The increase in $\sigp$ and $\sigz$ (second and third rows) in the barred simulations are also significantly larger than the unbarred simulations - especially within $R=0.5$~kpc. In general, the barred models at $t_2$ are more radially anisotropic than the unbarred models. This can be attributed to the dramatic increase in radial dispersion accompanied by only moderate increases in tangential and vertical dispersions.

Evidently the presence of the bar facilitates an increase in radial anisotropy during the growth of the CMC. This supports the idea that the elongated bar orbits are scattered by the CMC allowing the system as a whole to become rounder, without individual orbits becoming more tangential. In fact \citet{Shen04} showed that low energy bar supporting orbits are converted to rounder, chaotic orbits by the growth of a CMC.  In contrast in the unbarred systems, the adiabatic growth of a CMC induces a more tangentially biased velocity ellipsoid \citep{quinlan_etal_95} but angular momentum conservation limits the degree to which matter can flow inwards. 

We now see that the inverse correlation between $\Delta \sige/\sige_{\mathrm{init}}$ and inclination seen in Figure~\ref{fig:sige_smbh}~(right) can also be explained by considering Figure~\ref{fig:sig_beta}. In both types of models  $\sigz$ undergoes a significant increase due to the growth of the CMC. At low inclinations, $\sigz$ is the primary contributor to $\sige$, because the system is viewed more or less face-on. Thus the growth of a CMC produces a noticeable increase in $\sige$. However, at high inclinations, $\sige$ is dominated by $\sigr$ and $\sigp$, which, in the unbarred cases, are hardly affected by the growth of a CMC. As a result, the unbarred models (red/pink) show an inverse relationship between inclination and the change in $\sige$ between $t_1$ and $t_2$. This is also why in Figure~\ref{fig:sige_smbh}~(right) the barred disk (open blue squares) simulation showed a weaker dependence between $\Delta \sige/\sige_{\mathrm{init}}$ and inclination.

Interestingly, between $t_1$ and $t_2$,  $\betaz$ \emph{decreases} at small radii, even in the barred case. This can be attributed to the fact that the black hole scatters the low energy (radial) orbits, producing a more isotropic velocity ellipsoid \citep{Shen04}. Therefore a consequence of growing a CMC in a barred or unbarred galaxy is an overall decrease in $\betaz$ at small radii.
 
\subsection{Angular Momentum Transport}
\label{sec:ang_mom}

\citet{LyndenBell72} first showed that angular momentum in collisionless disks can be transferred outward via emission and absorption at the inner and outer Lindblad resonances. Several subsequent studies \citep{weinberg_85,Debattista00,Athanassoula03} showed that resonant material can exchange angular momentum between the bar and halo of a galaxy. Other recent studies \citep[e.g.,][]{Saha12} have investigated the transfer of angular momentum between the bar and bulge components. The exchange of angular momentum between morphological components of a galaxy has important implications for that galaxy's dynamical evolution. 

When a live halo is present, dynamical friction exerted by the halo on the bar can slow it down by allowing angular momentum exchange with the halo. It is important to note that in our simulations, which incorporate a static halo potential, a time-independent (nearly steady-state) bar is not expected to transfer significant amounts of angular momentum, since the torque exerted by such a bar on a star during one half of its orbit is of the same magnitude but opposite sign to the torque exerted on the second half of the orbit \citep{Binney08}. However when the potential of the bar is changing with time, as is the case when a central SMBH is growing, or if the bar strength or pattern speed are changing due to dynamical friction with the disk and bulge, a net transfer of angular momentum can result.
 
While an exhaustive discussion of the transfer of angular momentum from the inner to outer regions of our simulations is beyond the scope of this paper, we briefly consider how the presence of a bar influences such angular momentum exchange in our simulations and how this is related to the changes we saw in the measured $\sige$ and $\sigr$ profiles of barred galaxies, and the differences between barred and unbarred galaxies. In the preceding sections, we showed that changing the mass of the CMC by a factor of 10 alters the observed velocity dispersion by a mere 2\%. Therefore for the remainder of this paper we consider only the $10^8 \msol$ simulations, while examining the cause of the differences in the evolution of the barred and unbarred galaxies.

\begin{figure}
\centering{\includegraphics[scale=1.,width=.5\textwidth,trim=0.pt 0.pt 0.pt 0.pt,clip]{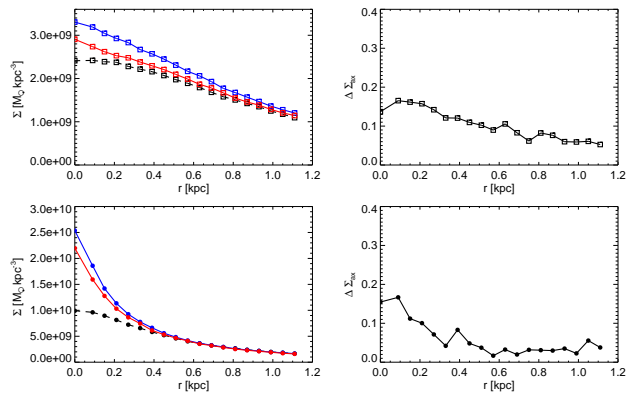}
\caption{Left: cylindrically averaged surface mass density profiles  as a function of cylindrical radius for initial model at $t_1$ (black), and after the growth of the CMC in the unbarred galaxy (red) and barred galaxy (blue) in disk-only model (top) and disk+bulge model bottom).  Right: Fractional difference in surface density in barred model relative to unbarred model $\Delta\Sigma_{\rm ax}$ as a function of radius for disk-only models (top) and for the disk+bulge models (bottom).}
\label{fig:mass_vs_r_comp}
}
\end{figure}

Figure~\ref{fig:mass_vs_r_comp} shows the cylindrically averaged mass density profiles as a function of radius for the initial models and for the barred and unbarred models after the growth of the CMC. The right hand panels plot the fractional difference in the surface mass density between the barred and unbarred models: $\Delta\Sigma_{\rm ax} = (\Sigma_{\rm barred} - \Sigma_{\rm ax})/\Sigma_{\rm ax} $ as a function of cylindrical radius. The increase in central mass surface density is between 5\% and 18\% higher in the barred galaxy than in unbarred galaxy (although the mass of the CMC is the same).

\begin{figure}
\centering{\includegraphics[scale=1.,width=.5\textwidth,trim=0.pt 0.pt 0.pt 0.pt,clip]{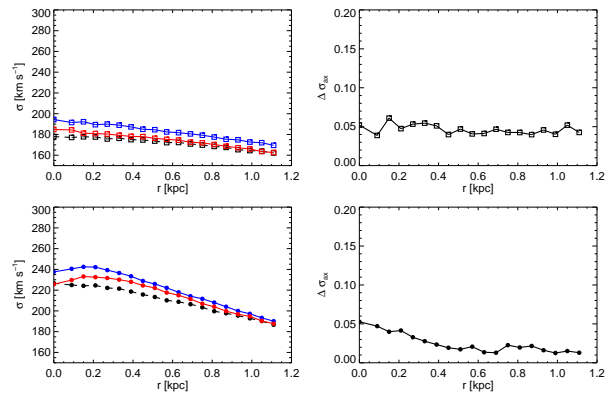}
\caption{Left: cylindrically averaged velocity dispersion profiles  as a function of cylindrical radius for initial model at $t_1$ (black),  after the growth of the CMC in the unbarred galaxy (red) and barred galaxy (blue) in disk-only model (top) and disk+bulge model bottom).  
Right:  Fractional difference in velocity dispersion $\Delta\sigma_{\rm ax}$ as a function of radius for disk-only models (top) and for the disk+bulge models (bottom).}
\label{fig:sig_vs_r_comp}
}
\end{figure}

Figure~\ref{fig:sig_vs_r_comp} shows the cylindrically averaged velocity dispersion profiles as a function of radius for the initial models and for the barred and unbarred models after the growth of the CMC. The right hand panels plot the fractional difference in the inrinsic velocity dispersion between the barred and unbarred models: $\Delta\sigma_{\rm ax} = (\sigma_{\rm barred} - \sigma_{\rm ax})/\sigma_{\rm ax} $ as a function of cylindrical radius. The increase in velocity dispersion is systematically higher by 5\%  in the barred galaxy in the disk-only case (top right panel) and between 2-5\% higher in the barred disk+bulge model (bottom right panel).

\begin{figure*}
\centering{\includegraphics[scale=1.,width=\textwidth,trim=0.pt 0.pt 0.pt 0.pt ,clip]{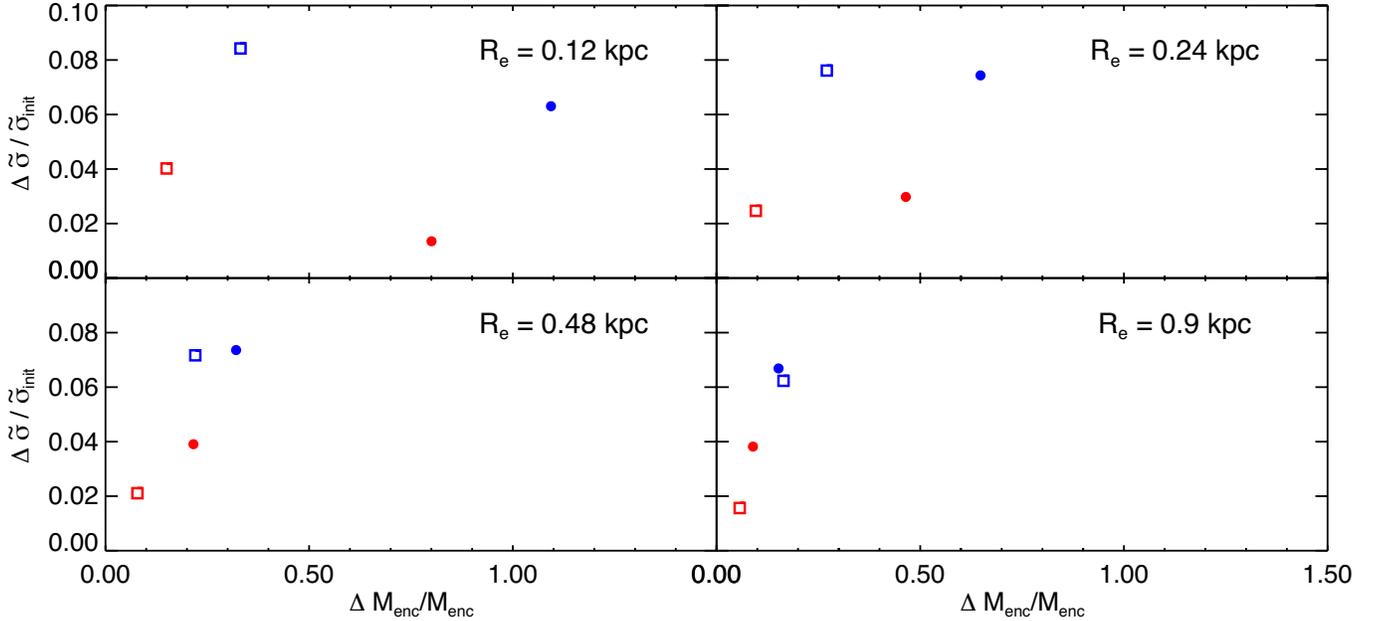}}
\caption{$\Delta \tilde\sige/\tilde\sigma_{\mathrm{init}}$ vs $\Delta M_{\mathrm{enc}}/M_{\mathrm{enc}}$ due to the growth of a CMC for the 4 different values of $\re$ used to measure $\sige$. Blue (red) denotes the presence (absence) of a bar. Squares (circles) denote disk (disk+bulge) simulations. For each value of $\re$, the barred cases have higher values of $\Delta \sige/\sigma_{init}$ and $\Delta M_{enc}/M_{enc}$ than their corresponding values in the unbarred case. Thus the presence of a bar results in a greater change in  both enclosed mass and stellar dispersion during the growth of a CMC.}
\label{fig:msig_cyl}
\end{figure*}

We define $\Delta\tilde\sigma/\tilde\sigma_{\mathrm{init}}$ as the fractional change (between $t_1$ and $t_2$) of the three dimensional intrinsic velocity dispersion $\tilde\sigma = \sqrt{(\sigr^2+\sigp^2+\sigz^2)}$, for all particles within the same cylindrical volume of radius $\re$. We define $\Delta M_{\mathrm{enc}}/M_{\mathrm{init}}$ as the fractional change (between $t_1$ and $t_2$)  in the mass enclosed by a specified cylindrical radius (note that $\Delta M$ excludes the mass contribution due to the CMC). In Figure~\ref{fig:msig_cyl} we plot $\Delta\tilde\sigma/\tilde\sigma_{\mathrm{init}}$ versus $\Delta M_{\mathrm{enc}}/M_{\mathrm{init}}$, for four different values of $\re$. Adding in the contribution of $\mbh$ would shift all the points towards the right, quite significantly for the smaller values of $\re$ ($M_{\mathrm{init}}\sim 10^8\msol \sim \mbh$ for $\re=0.12$) but cause only a small rightward shift for the larger values of $\re$ ($M_{\mathrm{init}}\sim 3\times 10^9\msol >> \mbh$ for $\re=0.9$).

A clear dichotomy exists between the barred (blue) and unbarred cases (red). At every value of $\re$ the barred models shows both a larger fractional increase in the enclosed mass and a larger fractional increase in the velocity dispersion of that. Thus the presence of a bar during the growth of a CMC facilitates both a higher mass increase within a specified radius and a higher 3-dimensional stellar velocity dispersion.  This is clear evidence that angular momentum transport in the barred simulations has facilitated the increase in both the enclosed mass and the velocity dispersion. 

It is also interesting to note that especially in the two smaller radial bins ($\re=0.12, 0.24$) although the increase in mass in the disk+bulge models (solid dots) is significantly larger than it is in the disk-only models (squares), the 3-dimensional velocity dispersion is larger in the disk-only models. Again this is due to the fact that in the disk-only case a larger fraction of the $x_1$ bar orbits survive the growth of the CMC, while in the disk+bulge case these orbits are more readily destroyed (most probably by the enhanced central density arising from the inflowing disk+bulge material). 

\begin{figure*}[t]

\centering{\includegraphics[scale=1.,width=0.8\textwidth,trim=0.pt 100.pt 0.pt 100.pt,clip]{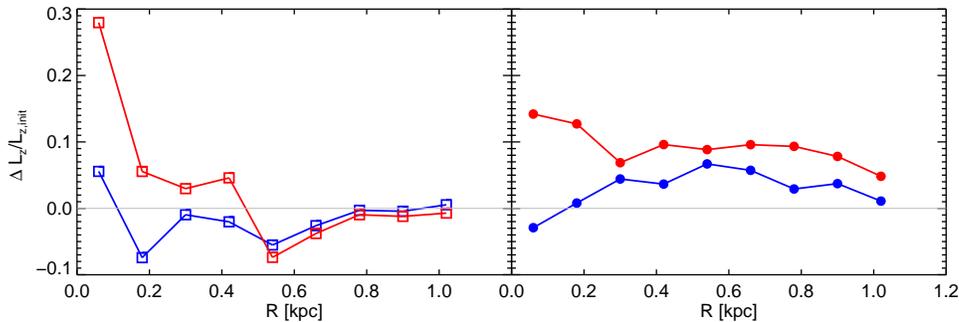}}

\caption{Fractional change in specific angular momentum in annular bins a function of radius for the disk-only simulations (left) and disk+bulge simulations (right). Blue (red) denotes barred (unbarred) simulations.}

\label{fig:ang_trans}

\end{figure*}

As final evidence for our claim that angular momentum transport by the bar plays a significant role in the velocity dispersion increase, in Figure~\ref{fig:ang_trans} we examine the fractional change in the average specific angular momentum of stars in the disk-only simulations (left) and disk+bulge simulations (right) for the barred (blue) and unbarred models at time $t_2$ relative to the value at $t_1$.  In the disk-only models it is clear that the change in the average specific angular momentum of stars in the barred systems is negative over most of the radial range plotted -- indicating that on average, stars have lost angular momentum. In contrast the corresponding unbarred system stars in the inner region have gained angular momentum at time $t_2$ relative to $t_1$, due to the adiabatic infall that gives rise to the growth of the central cusp that follows the growth of the CMC \citep{quinlan_etal_95}. Since this system has only weak spiral features incapable of transporting significant angular momentum, the specific angular momentum of stars has increased as the cusp formed.

Recall from Figure~\ref{fig:msig_cyl} that at each radius the fractional increase in enclosed mass  $\Delta M_{\mathrm{enc}}/M_{\mathrm{init}}$ within each cylindrical radial bin is always larger in the barred system than for the corresponding unbarred system. 
In a collisionless simulation the net angular momentum of the system is conserved. If significant angular momentum transport does not occur, then an increase in specific angular momentum is expected as matter is drawn inwards. The fact that the angular momentum per particle in most of the inner 0.8~kpc of the barred  galaxy has {\it decreased} shows that some of the angular momentum must have been transported outwards by the bar.  In the disk+bulge models (right)  we see that change in the specific angular momentum of the barred galaxy (blue) is always smaller than for the unbarred galaxy also pointing to outward transport. 
 

Thus Figures~\ref{fig:mass_vs_r_comp}, \ref{fig:sig_vs_r_comp}, \ref{fig:msig_cyl}, and \ref{fig:ang_trans} clearly demonstrate that the time dependent bar-potential resulting from the growing SMBH results in angular momentum transport that is responsible for increasing the central mass of stars and the radial anisotropy of orbits.

\subsection{Effects of Bar Kinematics on SMBH Mass Measurement}
\label{sec:measure_mbh}

\begin{figure*}

\centering{\includegraphics[scale=1.,height=.3\textheight,trim=0.pt 0.pt 0.pt 125.pt,clip]{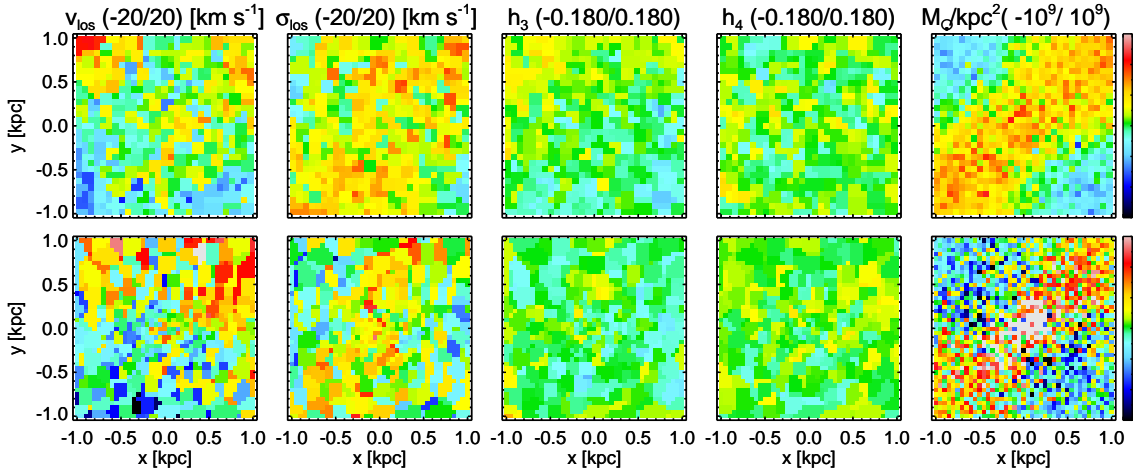}}

\caption{Maps showing the difference between quantities (from left to right: $\vlos$, $\siglos$, $h_3$, $h_4$ and surface mass density) measured in the barred simulations at $t_2$ and the same quantity measured in its unbarred counterpart at $t_2$. Top row shows the difference maps for the pure-disk simulations. Bottom row shows the difference maps for the disk+bulge simulations.}

\label{fig:diff_bar-ax}
\end{figure*}

The measured values of $\mbh$ in barred galaxies compiled by \citet{Graham11} come from a variety of dynamical measurement techniques: gas kinematics, stellar dynamics, and reverberation mapping. In the most recent compilation of galaxies with dynamically measured SMBH masses \citep{mcconnell_ma_13} consisting of 72 galaxies, nearly 50\% of the SMBH mass measurements are derived via stellar dynamical methods. Stellar dynamical methods entail modeling the nuclear stellar kinematics either via the technique referred to as the Schwarzschild orbit superposition method \citep[e.g][]{schwarzschild_79,vandermarel_etal_98,cretton_etal_99,gebhardt_etal_03,valluri_etal_04,vandenbosch_etal_08} or by solving the axisymmetric  Jeans equations  \citep{Binney08,cappellari_08}. Both methods simultaneously optimize the fit to the 3-dimensional mass distribution (including the mass of the unknown SMBH), the surface brightness distribution and the observed LOSVDs to constrain the best fit values of $\mbh$ and the M/L ratio of the stars. A small number of elliptical galaxies in this sample have been modeled with a (non-rotating) triaxial orbit superposition code \citep[e.g.][]{vdBosch10}, but most stellar dynamical $\mbh$ measurements have been made with axisymmetric modeling codes. 

We examined the table of 72 galaxies with dynamically measured SMBH presented by \citep{mcconnell_ma_13} and find that of the sample of $\sim 35$ galaxies in which $\mbh$ was measured via stellar dynamical methods, 17 are S0 or spiral galaxies. Of these, 5 galaxies (29\%) are classified as barred in the NASA Extragalactic Database\footnote{http://ned.ipac.caltech.edu}, and another 6 (35\%) are edge-on galaxies in which a bar would be difficult to detect should it exist.  We note that the total fraction of barred + edge-on galaxies (64\%) is comparable to the fraction of local disk galaxies that contain bars \citep[e.g.][]{Knapen99,eskridge_etal_00,menendez_etal_07,marinova_jogee_07,sheth_etal_08}. The black hole masses for {\em all} these galaxies have been obtained using axisymmetric stellar dynamical codes. In this section we examine qualitatively the possible systematic biases that the assumption of axisymmetry might have on the measured mass of the SMBH in a barred galaxy. We defer a more quantitative study to a future paper.

The process of measuring the dynamical mass of an SMBH from the kinematics of stars in the nucleus suffers from the well known mass-anisotropy degeneracy \citep{binney_mamon_82}. In this classic paper the authors showed that the degeneracy arises because orbits of stars in elliptical galaxies and the bulges of disk galaxies can have a wide range of possible velocity anisotropy distributions. A large line-of-sight central  stellar velocity dispersion in the nucleus could be the result of  a large central SMBH about which stars move on primarily tangential orbits, or could equally well be the result of stars on primarily radial orbits moving around a much smaller (or no) central SMBH. In axisymmetric and spherical models the degeneracy between mass and velocity anisotropy can be lifted by the use of information contained in the shapes of the stellar LOSVDs.  It is customary to use Gauss-Hermite coefficients to represent the deviations of an LOSVD from a Gaussian shape \citep{VDM93,gerhard_93}. In axisymmetric or spherical systems, stars on predominantly radial orbits will give rise to LOSVDs with positive $h_4$ parameter, while stars on predominantly tangential orbits produce LOSVDs with  negative $h_4$ parameters. An isotropic velocity distribution will produce an LOSVD with $h_4\sim 0$. Degeneracy between mass and anisotropy is lifted by ensuring that the orbit superposition method simultaneously fits at least  $\siglos$, and $h_4$.

In the immediate vicinity of a SMBH, the presence of a large fraction of stars at high velocities causes an increase in the amplitudes of the high velocity wings of the LOSVD resulting in large positive values of $h_4$ \citep{vandermarel94}. These high velocity tails provide strong constraints on kinematics of stars in the vicinity of the SMBH and on its mass.

Currently there are no stellar dynamical modeling codes that are can measure the masses of SMBHs in barred galaxies. However, recently \citet{lablanche_etal_12} used an axisymmetric stellar dynamical modeling code  \citep[Jeans Anisotropic MGE (JAM) method,][]{cappellari_08} to assess the accuracy with which the stellar $M/L$ ratio and intrinsic velocity anisotropy could be recovered. They applied the method to a sample of $N$-body simulations of barred S0 galaxies and showed that biases in the determination of M/L primarily arise due to the application of an axisymmetric modeling code to a barred galaxy. They find that for $\philon=45^\circ$ and $i>30$ the measured stellar $M/L$ ratio is essentially unbiased, but errors of up to 15\% can arise due to varying orientation of the bar and inclination of the disk.  Furthermore they find that when a bar is present, the inferred velocity anisotropy can be significantly in error. 

While it is beyond the scope of this paper to carry out a similar exercise to assess the systematic biases that would be introduced into the measured masses of SMBHs by using axisymmetric stellar dynamical codes, we will qualitatively examine the nature and the direction of the bias.

Figure~\ref{fig:diff_bar-ax} shows the stellar kinematic difference maps in the inner $\pm1$~kpc region for the pure disk simulations (top row) and the  disk+bulge simulations (bottom row). The maps show the differences in the kinematic quantities $\vlos, \siglos, h_3, h_4$ and the projected mass density $\Sigma$. In each panel we plot the difference in a specific quantity between barred and unbarred models after the growth of the SMBH in each pixel in the field of view. The angle of inclination of the disk and $\philon$ are both set to 45$^\circ$. Pixels that are colored green  indicate that there is no difference between the barred and unbarred models; yellow and red pixels imply that the quantity in the barred galaxy ($\vlos, \siglos, h_3, h_4, \Sigma$) is higher and blue pixels indicate that the quantity in the barred galaxy is lower than it is in its unbarred counterpart. The values in parenthesis above each panel indicate the range of the difference in the quantities. Notice that the difference maps show that $\siglos$ in the inner regions of the map is always red/yellow indicating that it is systematically higher in the barred models than in the unbarred models (see also Fig.~\ref{fig:sig_vs_r_comp}) while $h_4$ is green/blue signifying that it is generally lower than in the unbarred models.  

Therefore, if a barred galaxy is modeled with the assumption of axisymmetry, the dynamical model will attempt to fit  the negative $h_4$ by putting a large fraction of orbits on tangential orbits, while the requirement to simultaneously fit a large $\siglos$ would require a larger enclosed mass than than one would infer from the same mass distribution in an unbarred model. The standard approach in stellar dynamical modeling is to hold the M/L ratio of the galaxy fixed \citep[however see][]{mcconnell_etal_13_mbyl}. When the M/L ratio is held fixed it is largely determined by kinematic constraints outside the sphere-of-influence of the SMBH, and the M/L ratio of stars in inner part of the bulge is likely to be underestimated.  Since bar-induce evolution can significantly increase mass inflow from large radii to small radii, the standard practice of holding M/L fixed will also result in $\mbh$ being overestimated. A striking example of this is seen in NGC~4151 which has recently been modeled by \citet{Onken13}.

We also examined the LOSVDs of stars within the sphere-of-influence of the CMC ($r_s$) in both the barred and unbarred models (for $i=45, \philon=45$). We find that although $h_4$ is negative on average within $R_e$, within $r_s$ -- the region where the CMC dominates the dynamics of stars  -- LOSVDs of the barred galaxies in our simulations have  30\%-50\% larger values of $h_4$ than the corresponding unbarred galaxies (for the same mass of CMC). This is due to a combination of the increased radial velocity anisotropy of stars resulting from the growth of the CMC (seen in Fig.~\ref{fig:sig_beta}) and streaming motions along the bar. Since the $\siglos$ values are also about 5\% larger in the barred models than in the unbarred models, we predict that even if the sphere-of-influence of the SMBH is resolved, the anisotropic velocity distribution will also result in an over-estimate of $\mbh$. The idea that the high central velocity dispersions of nearly end-on bars can be mistaken for  central black holes is not new \citep{gerhard88} and has recently been invoked as an alternative explanation \citep{emsellem13} for the claimed over-massive black hole in NGC~1277 \citep{vandenbosch_etal_12}.

Using unbarred dynamical modeling codes to measure the masses of SMBHs in barred galaxies is therefore likely to result in a systematic overestimate of $\mbh$, regardless of whether the sphere-of-influence of the SMBH is resolved or not.  Since the fraction of barred galaxies with stellar dynamical determinations of $\mbh$ is currently quite a small fraction of all the SMBH measurements used in the most recent $\msigma$ relation, this is unlikely to significantly alter this relation or offset of barred galaxies from it. However, the effect of using unbarred models to measure the mass of SMBHs should be examined quantitatively in the future.

\section{SUMMARY}
\label{sec:summary}

We simulated the growth of CMCs  representing SMBHs (with mass up to 0.2\% of the mass of the disk) in $N$-body simulations of disk galaxies both with and without bars and both with and without  classical bulges. Our main findings are 
\begin{itemize}
\item The growth of a CMC in a barred galaxy produces an increase in $\sigma$ that is $\sim5-8\%$ larger than in an axisymmetric counterpart.
\item The measured value of $\sige$ is relatively insensitive to the choice of $\re$.
\item Orientation effects are only partially responsible for the different measurements of $\sige$ obtained from barred and unbarred galaxies.
\item The growth of a CMC alters the potential of the bar, enabling  outward transport of angular momentum and a consequent increase in the central mass of stars. The increase in central mass is partly responsible for the increase in central velocity dispersion.
\item The change in $\sigma$ and $\Delta\sigma_{\rm ax}$ is fairly insensitive to an order of magnitude change in $\mcmc$, showing that it is the evolution of the bar potential induced by CMC growth, rather than the final mass of the CMC, that is the primary factor driving the increase in $\sigma$.  
\item The scattering of bar orbits by the central CMC results in an increase in all components of the velocity dispersion, but particularly the radial velocity dispersion. In contrast CMC growth in an axisymmetric disk induces an tangentially biased velocity dispersion. Thus a strong radial anisotropy and a large offset in $\sigma$ are likely to be predictors of bar induced CMC growth.
\item We predict an over-estimate of $\mbh$ if axisymmetric stellar dynamical modeling codes are used to measure the masses of SMBHs in barred galaxies, especially if M/L ratios are assumed to be independent of radius.
\end{itemize}

\section{DISCUSSION and CONCLUSIONS}
\label{sec:discussion}

We have  investigated the effect that the adiabatic growth of a CMC representing an SMBH has on the nuclear kinematics in galaxies with pre-existing bars. We compared these barred simulations to unbarred analogues with identical radially averaged mass and velocity distributions which were constructed by scrambling the barred disk particles in azimuthal angle. In these simulations we have assumed that the galaxy's disk/bulge and bar are fully formed before the growth of the SMBH begins, and our focus is on the effect that this SMBH has on the system. Clearly this is a simplification of reality but it allows us to isolate the effects of various observing conditions from the dynamical effects of growing an SMBH. We do not consider the possibility that a disk with a pre-existing bulge and SMBH may become unstable to bar formation, which would also alter the observed kinematics, since this scenario is considered by \citet{Hartmann13}. This latter work  shows that disk heating and angular momentum transport due to bar formation may be a key contributor to the increased dispersion of barred galaxies. 

AGN feedback and gas dynamics have been ignored here and both can have  important effects on the dynamics of the host galaxy.  AGN feedback may couple the SMBH to its host, since only a small fraction of the energy available via accretion processes is required to significantly alter the kinematics and evolution of the host galaxy \citep{Silk98,Fabian99,DiMatteo05}. These works show that feedback from SMBH accretion may strip the host galaxy of gas, thus halting both star formation and SMBH growth, leading to the black hole scaling relations we observe today. However, \citet{Alcazar13} show that self-regulating feedback due to the growth of an SMBH is not required to produce the observed galaxy black hole--galaxy scaling relations. Instead, gravitational torques \citep[e.g.][]{Hopkins11} could limit accretion, ultimately allowing for the rapid growth of young SMBHs.  This is an active area of current research and at present it is not clear whether SMBH feedback, gravitational instabilities, or some other mechanism is driving the observed black hole scaling relations.

It is important to recognize that the growth mechanism of SMBHs in morphologically different galaxies need not be the same. For instance, SMBHs with masses $\mbh \sim 10^9 \msol$ typically found in massive elliptical galaxies have probably grown via hierarchical merging accompanied by rapid accretion whereas the SMBHs with masses of $\mbh \sim 10^7 \msol$ residing in disk galaxies may have grown primarily via secular accretion processes. Recent HST WFC3/Infrared imaging observations of heavily dust obscured AGN at redshifts $z \sim 1 - 3$ find that almost 90\% of the host galaxies are disks \citep{Schawinski12} suggesting that significant growth of SMBHs could be occurring via secular processes in disks rather than in major merger events. In fact, multiwavelength studies of AGN from $z\sim0-3$ show that only the most luminous AGN hosts are ellipticals also suggesting that a significant fraction of SMBH growth occurs in disk galaxies \citep{treister_etal_12}. \citet{cisternas_etal_11} find little evolution in the $\mbh$-host stellar mass relation since $z\sim 0.9$. However, since a significant fraction of the galaxies at higher redshifts have a prominent disk component, their bulges are undermassive. They argue that over the last 7~Gyr there must have been a  redistribution of stellar mass from the disk to the bulge, perhaps driven by secular evolution.  The influence of the bar on the growth of the bulge mass and bulge velocity dispersion demonstrated in this paper is one secular evolution mechanism that could have played a role in this redistribution. Although the precise criteria for distinguishing between pseudo-bulges and classical bulges have been the subject of debate for nearly a decade \citep[for reviews of the status see][]{Kormendy04,Graham12}, pseudo-bulges are generally thought to have formed as a result of secular evolution in a disk galaxy (e.g. due to outward transport of angular momentum, and inward flow of matter resulting from a time-varying bar potential; for a recent review, see \citealt{Athanassoula12}). In contrast, classical bulges are thought to have formed via mergers.  Clearly the issue of whether or not there is clear observational evidence for differences in the $\msigma$ relationship based on morphological type is an issue that is still in a state of flux.

 Recently \citet{debattista_etal_13} showed that if disks reform and grow around bulges with a pre-existing SMBH, the velocity dispersion of the bulge itself can increase due to adiabatic compression by the disk, requiring the SMBH to grow by 50-60\% just to stay on the $\msigma$ relation. Thus the small observed scatter in the BH-host galaxy scaling relations suggest strongly that  BHs ``know about'' their hosts. \citet{Hopkins09} argue that the amount of gas that formed stars in the spheroid of host galaxies shows an order-of-magnitude scatter and that unless black hole growth is self-regulated via feedback processes, the scatter in BH-scaling relations would be significantly larger than is observed. Searching for galaxies of specific morphological types which show systematic deviations from scaling relations which may arise due to secular evolution allows us to confirm or reject the idea of tightly self-regulated SMBH growth. 
 
In Section~\ref{sec:intro} we noted that barred galaxies lie 0.3 dex below the $\msigma$ relation defined by unbarred galaxies but do not appear to be offset from the $M$--$L$ relation. \citet{Graham08a}  has used this to argue  that $\mbh$ values in barred galaxies are not under massive relative to unbarred galaxies. If neglecting bar kinematics in stellar dynamical modeling can result in an overestimate of $\mbh$ as argued above, their true values could be even lower than their currently estimated values. 
Since local samples of late type galaxies show that nearly 65\% of them are barred \citep[e.g.][]{Knapen99,Eskridge00,Sheth08} and since late type galaxies may contain the vast majority of black holes below $5\times10^7\msol$ \citep[e.g.][]{Graham11}, there is a need for dynamical modeling methods that can measure the masses of SMBH in barred galaxies. 

The dark matter halos in the disk galaxies in our simulations  were embedded in static (rigid) dark matter halos with shallow central density cores. Previous work has shown that the presence of a live dark matter halo, especially one with a steep central density cusp \citep[e.g.][]{Debattista00,Athanassoula03}, can slow down the pattern speed of a bar due to dynamical friction which causes energy and angular momentum of the bar to be lost to the halo. In general, the presence of a live halo enhances the process of angular momentum transport from the bar to the halo, the details of which depend on the distribution function of the halo. Previous studies suggest that with a live halo, the amount of matter that flows inward could be somewhat larger than in the simulation presented here; this might cause an even greater increase in $\sige$ than we obtained. Further studies of SMBH growth in disk galaxies with live halos are necessary to quantify the extent of the increase in such simulations. 

It is clear that the effect of the SMBH on $\sige$ arises from the effect of the growing SMBH potential on the interaction between the bar and the disk (disk+bulge). The SMBH alters the observable $\sige$ (within the effective radius) well outside the sphere of influence of the SMBH. Hence, the  presence of a bar during the growth of an SMBH may partially explain the rightward offset of barred galaxies from the $\msigma$ relation defined by unbarred galaxies presented in \citet{Graham11}. 

\acknowledgements  MV and JB are supported by funds from the US National Science Foundation under grant AST-09AST-0908346.  MV also acknowledges support from University of Michigan's Elizabeth Crosby grant.  JS is supported in part by the National Natural Science Foundation of China under grant No. 11073037, by 973 Program of China under grant No. 2009CB824800, and by the CAS Bairen Grant. V.P.D. is supported in part by UK STFC Consolidated grant \# ST/J001341/1. MV and VPD thank the Aspen Center for Physics and NSF Grant \#1066293 for hospitality during the final stages of this project. MV would like to thank Yijia Tang for her important contributions in the early stages of this project, and E. Athanassoula for useful discussions. We thank M. Cappellari for making publicly available the Voronoi binning routines used in this paper. This work has made use of  NASA's Astrophysics Data System Bibliographic Services (http://adsabs.harvard.edu/NASA ADS).

\bibliography{barred_offset}
\end{document}